\documentclass{amsart}
\usepackage{amsmath}
\usepackage{amsfonts}
\usepackage{amssymb}
\usepackage{graphicx}
\newtheorem{theorem}{Theorem}

\newtheorem{proposition}[theorem]{Proposition}

\theoremstyle{remark}

\newtheorem{example}[theorem]{Example}
\numberwithin{equation}{section}   
\numberwithin{theorem}{section}

\let\frak = \mathfrak

\begin{document}
\title[Geometric quantization and Segal--Bargmann transform]{Geometric quantization and the generalized Segal--Bargmann transform for Lie
groups of compact type}
\author{Brian C. Hall}
\address{Department of Mathematics\\
University of Notre Dame \\
Notre Dame, IN 46556, U.S.A. }
\email{bhall{@}nd.edu}
\begin{abstract}
Let $K$ be a connected Lie group of compact type and let $T^{\ast }(K)$ be
its cotangent bundle. This paper considers geometric quantization of $%
T^{\ast }(K),$ first using the vertical polarization and then using a
natural K\"{a}hler polarization obtained by identifying $T^{\ast }(K)$ with
the complexified group $K_{\mathbb{C}}.$ The first main result is that the
Hilbert space obtained by using the K\"{a}hler polarization is naturally
identifiable with the generalized Segal--Bargmann space introduced by the
author from a different point of view, namely that of heat kernels. The
second main result is that the pairing map of geometric quantization
coincides with the generalized Segal--Bargmann transform introduced by the
author. This means that the pairing map, in this case, is a constant
multiple of a unitary map. For both results it is essential that the
half-form correction be included when using the K\"{a}hler polarization.

These results should be understood in the context of results of K. Wren and
of the author with B. Driver concerning the quantization of $(1+1)$%
-dimensional Yang--Mills theory. Together with those results the present
paper may be seen as an instance of ``quantization commuting with
reduction.''
\end{abstract}
\maketitle
\setcounter{tocdepth}{1}
\tableofcontents

\section{Introduction}

The purpose of this paper is to show how the generalized Segal--Bargmann
transform introduced by the author in \cite{H1} fits into the theory of
geometric quantization. I begin this introduction with an overview of the
generalized Segal--Bargmann transform and its applications. I continue with a
brief description of geometric quantization and I conclude with an outline of
the results of this paper. The reader may wish to begin with Section 5, which
explains how the results work out in the $\mathbb{R}^{n}$ case.

\subsection{The generalized Segal--Bargmann transform}

See the survey paper \cite{H7} for a summary of the generalized
Segal--Bargmann transform and related results.

Consider a classical system whose configuration space is a connected Lie group
$K$ of compact type. Lie groups of compact type include all compact Lie
groups, the Euclidean spaces $\mathbb{R}^{n},$ and products of the two (and no
others--see Section 7). As a simple example, consider a rigid body in
$\mathbb{R}^{3},$ whose rotational degrees of freedom are described by a
system whose configuration space is the compact group $\mathrm{SO}(3).$

For a system whose configuration space is the group $K,$ the corresponding
phase space is the cotangent bundle $T^{\ast}(K).$ There is a natural way to
identify $T^{\ast}(K)$ with the \textit{complexification} $K_{\mathbb{C}} $ of
$K.$ Here $K_{\mathbb{C}}$ is a certain connected complex Lie group whose Lie
algebra is the complexification of $Lie(K)$ and which contains $K$ as a
subgroup. For example, if $K=\mathbb{R}^{n}$ then $K_{\mathbb{C}}%
=\mathbb{C}^{n}$ and if $K=\mathrm{SU}(n)$ then $K_{\mathbb{C}}=\mathrm{SL}%
(n;\mathbb{C}).$

The paper \cite{H1} constructs a generalized Segal--Bargmann transform for
$K.$ (More precisely, \cite{H1} treats the compact case; the $\mathbb{R}^{n}$
case is just the classical Segal--Bargmann transform, apart from minor
differences of normalization.) The transform is a unitary map $C_{\hbar}$ of
$L^{2}(K,dx)$ onto $\mathcal{H}L^{2}(K_{\mathbb{C}},\nu_{\hbar}(g)\,dg), $
where $dx$ and $dg$ are the Haar measures on $K$ and $K_{\mathbb{C}},$
respectively, and where $\nu_{\hbar}$ is the $K$-invariant \textit{heat
kernel} on $K_{\mathbb{C}}.$ Here $\hbar$ is Planck's constant, which is a
parameter in the construction (denoted $t$ in \cite{H1}). The transform itself
is given by
\[
C_{\hbar}f=\text{ analytic continuation of }e^{\hbar\Delta_{K}/2}f,
\]
where the analytic continuation is from $K$ to $K_{\mathbb{C}}$ with $\hbar$
fixed. The results of the present paper and of \cite{Wr} and \cite{DH} give
other ways of thinking about the definition of this transform. (See below and
Section 3 for a discussion of \cite{Wr,DH}.)

The results of \cite{H1} can also be formulated in terms of coherent states
and a resolution of the identity, as described in \cite{H1} and in much
greater detail in \cite{HM}. The isometricity of the transform and the
resolution of the identity for the coherent states are just two different ways
of expressing the same mathematical result.

The results of \cite{H1} extend to systems whose configuration space is a
compact homogeneous space, such as a sphere, as shown in \cite[Sect. 11]{H1}
and \cite{St}. However the group case is special both mathematically and for
applications to gauge theories. In particular the results of the present paper
do \textit{not} extend to the case of compact homogeneous spaces.

The generalized Segal--Bargmann transform has been applied to the Ashtekar
approach to quantum gravity in \cite{A}, as a way to deal with the ``reality
conditions'' in the original version of this theory, formulated in terms of
complex-valued connections. (See also \cite{Lo}.)

More recently progress has been made in developing a purely real-valued
version of the Ashtekar approach, using compact gauge groups. In a series of
six papers (beginning with \cite{T2}) T. Thiemann has given in this setting a
diffeomorphism-invariant construction of the Hamiltonian constraint, thus
giving a mathematically consistent formulation of quantum gravity. In an
attempt to determine whether this construction has ordinary general relativity
as its classical limit, Thiemann and co-authors have embarked on a program
\cite{T3,TW1,TW2,TW3,STW} to construct coherent states that might approximate
a solution to classical general relativity. These are to be obtained by gluing
together the coherent states of \cite{H1} for a possibly infinite number of
edges in the Ashtekar scheme. This program requires among other things a
detailed understanding of the properties of the coherent states of \cite{H1}
for one fixed compact group $K,$ which has been worked out in the case
$K=\mathrm{SU}(2)$ in \cite{TW1}.

In another direction, K. K. Wren \cite{Wr}, using a method proposed by N. P.
Landsman \cite{La1}, has shown how the coherent states of \cite{H1} arise
naturally in the canonical quantization of $(1+1)$-dimensional Yang--Mills
theory on a spacetime cylinder. The way this works is as follows. (See Section
3 for a more detailed explanation.) For the canonical quantization of
Yang--Mills on cylinder, one has an infinite-dimensional ``unreduced''
configuration space consisting of $K$-valued connections over the spatial
circle, where $K$ is the structure group. One is then supposed to pass to the
``reduced'' or ``physical'' configuration space consisting of connections
modulo gauge transformations. It is convenient to work at first with ``based''
gauge transformations, those equal to the identity at one fixed point in the
spatial circle. In that case the reduced configuration space, consisting of
connections modulo based gauge transformations over $S^{1},$ is simply the
structure group $K.$ (This is because the one and only quantity invariant
under based gauge transformations is the holonomy around the spatial circle.)

Wren considers the ordinary ``canonical'' coherent states for the space of
connections and then ``projects'' these (using a suitable regularization
procedure) onto the gauge-invariant subspace. The remarkable result is that
after projection the ordinary coherent states for the space of connections
become precisely the generalized coherent states for $K$, as originally
defined in \cite{H1}. Wren's result was elaborated on by Driver--Hall
\cite{DH} and Hall \cite{H8}, in a way that emphasizes the Segal--Bargmann
transform and uses a different regularization scheme. These results raise
interesting questions about how geometric quantization behaves under
reduction--see Section 3.

Finally, as mentioned above, we can think of the Segal--Bargmann transform for
$K$ as a resolution of the identity for the corresponding coherent states. The
coherent states then ``descend'' to give coherent states for any system whose
configuration space is a compact homogeneous space \cite[Sect. 11]{H1},
\cite{St}. Looked at this way, the results of \cite{H1,St} fit into the large
body of results in the mathematical physics literature on generalized coherent
states. It is very natural to try to construct coherent states for systems
whose configuration space is a homogeneous space, and there have been previous
constructions, notably by C. Isham and J. Klauder \cite{IK} and De Bi\`{e}vre
\cite{De}. However, these constructions, which are based on extensions of the
Perelomov \cite{P} approach, are \textit{not} equivalent to the coherent
states of \cite{H1,St}. In particular the coherent states of \cite{IK} and
\cite{De} do not in any sense depend holomorphically on the parameters, in
contrast to those of \cite{H1,St}.

More recently, the coherent states of Hall--Stenzel for the case of a 2-sphere
were independently re-discovered, from a substantially different point of
view, by K. Kowalski and J. Rembieli\'{n}ski \cite{KR1}. (See also
\cite{KR2}.) The forthcoming paper \cite{HM} explains in detail the coherent
state viewpoint, taking into account the new perspectives offered by Kowalski
and Rembieli\'{n}ski \cite{KR1} and Thiemann \cite{T1}. In the group case, the
present paper shows that the coherent states of \cite{H1} can be obtained by
means of geometric quantization and are thus of ``Rawnsley type''
\cite{Ra1,RCG}.

\subsection{Geometric quantization}

A standard example in geometric quantization is to show how the
Segal--Bargmann transform for $\mathbb{R}^{n}$ can be obtained by means of
this theory. Furthermore, the standard method for constructing other
Segal--Bargmann-type Hilbert spaces of holomorphic functions (and the
associated coherent states) is by means of geometric quantization. Since
\cite{H1} is not formulated in terms of geometric quantization, it is natural
to apply geometric quantization in that setting and see how the results
compare. A first attempt at this was made in \cite[Sect. 7]{H4}, which used
``plain'' geometric quantization and found that the results were not
equivalent to those of \cite{H1}. The present paper uses geometric
quantization with the ``half-form correction'' and the conclusion is that
geometric quantization with the half-form correction \textit{does} give the
same results as \cite{H1}. In this subsection I give a brief overview of
geometric quantization, and in the next subsection I summarize how it works
out in the particular case at hand. See also Section 5 for how all this works
in the standard $\mathbb{R}^{n}$ case.

For quantum mechanics of a particle moving in $\mathbb{R}^{n}$ there are
several different ways of expressing the quantum Hilbert space, including the
position Hilbert space (or Schr\"{o}dinger representation) and the
Segal--Bargmann (or Bargmann, or Bargmann-Fock) space. The position Hilbert
space is $L^{2}(\mathbb{R}^{n}),$ with $\mathbb{R}^{n}$ thought of as the
position variables. The Segal--Bargmann space is the space of holomorphic
functions on $\mathbb{C}^{n}$ that are square-integrable with respect to a
Gaussian measure, where $\mathbb{C}^{n}=\mathbb{R}^{2n}$ is the phase space.
(There are also the momentum Hilbert space and the Fock symmetric tensor
space, which will not be discussed in this paper.) There is a natural unitary
map that relates the position Hilbert space to the Segal--Bargmann space,
namely the Segal--Bargmann transform.

One way to understand these constructions is in terms of geometric
quantization. (See Section 5.) In geometric quantization one first constructs
a pre-quantum Hilbert space over the phase space $\mathbb{R}^{2n}. $ The
prequantum Hilbert space is essentially just $L^{2}\left(  \mathbb{R}%
^{2n}\right)  .$ It is generally accepted that this Hilbert space is ``too
big''; for example, the space of position and momentum operators does not act
irreducibly. To get an appropriate Hilbert space one chooses a
``polarization,'' that is (roughly) a choice of $n$ out of the $2n$ variables
on $\mathbb{R}^{2n}.$ The quantum Hilbert space is then the space of elements
of the prequantum Hilbert space that are independent of the chosen $n$
variables. So in the ``vertical polarization'' one considers functions that
are independent of the momentum variables, hence functions of the position
only. In this case the quantum Hilbert space is just the position Hilbert
space $L^{2}\left(  \mathbb{R}^{n}\right)  .$ Alternatively, one may identify
$\mathbb{R}^{2n}$ with $\mathbb{C}^{n}$ and consider complex variables
$z_{1},\cdots,z_{n},$ and $\bar{z}_{1},\cdots,\bar{z}_{n}.$ The Hilbert space
is then the space of functions that are ``independent of the $\bar{z}_{k}%
$'s,'' that is, holomorphic. In this case the quantum Hilbert space is the
Segal--Bargmann space.

More precisely, the prequantum Hilbert space for a symplectic manifold
$\left(  M,\omega\right)  $ is the space of sections of a
line-bundle-with-connection $L$ over $M,$ where the curvature of $L$ is given
by the symplectic form $\omega.$ A real polarization for $M$ is a foliation of
$M$ into Lagrangian submanifolds. A K\"{a}hler polarization is a choice of a
complex structure on $M$ that is compatible with the symplectic structure, in
such a way that $M$ becomes a K\"{a}hler manifold. The quantum Hilbert space
is then the space of sections that are covariantly constant along the leaves
of the foliation (for a real polarization) or covariantly constant in the
$\bar{z}$-directions (for a complex polarization). Since the leaves of a real
polarization are required to be Lagrangian, the curvature of $L$ (given by
$\omega$) vanishes along the leaves and so there exist, at least locally,
polarized sections. Similarly, the compatibility condition between the complex
structure and the symplectic structure in a complex polarization guarantees
the existence, at least locally, of polarized sections.

A further ingredient is the introduction of ``half-forms,'' which is a
technical necessity in the case of the vertical polarization and which can be
useful even for a K\"{a}hler polarization. The inclusion of half-forms in the
K\"{a}hler-polarized Hilbert space is essential to the results of this paper.

If one has two different polarizations on the same manifold then one gets two
different quantum Hilbert spaces. Geometric quantization gives a canonical way
of constructing a map between these two spaces, called the pairing map. The
pairing map is not unitary in general, but it is unitary in the case of the
vertical and K\"{a}hler polarizations on $\mathbb{R}^{2n}$. In the
$\mathbb{R}^{2n}$ case, this unitarity can be explained by the Stone-von
Neumann theorem. I do the calculations for the $\mathbb{R}^{2n}$ case in
Section 5; the reader may wish to begin with that section.

Besides the $\mathbb{R}^{2n}$ case, there have not been many examples where
pairing maps have been studied in detail. In particular, the only works I know
of that address unitarity of the pairing map outside of $\mathbb{R}^{2n} $ are
those of J. Rawnsley \cite{Ra2} and K. Furutani and S. Yoshizawa \cite{FY}.
Rawnsley considers the cotangent bundle of spheres, with the vertical
polarization and also a certain K\"{a}hler polarization. Furutani and
Yoshizawa consider a similar construction on the cotangent bundle of complex
and quaternionic projective spaces. In these cases the pairing map is not
unitary (nor a constant multiple of a unitary map).

\subsection{Geometric quantization and the Segal--Bargmann transform}

An interesting class of symplectic manifolds having two different natural
polarizations is the following. Let $X$ be a real-analytic Riemannian manifold
and let $M=T^{\ast}(X).$ Then $M$ has a natural symplectic structure and a
natural vertical polarization, in which the leaves of the Lagrangian foliation
are the fibers of $T^{\ast}\left(  X\right)  .$ By a construction of Guillemin
and Stenzel \cite{GStenz1,GStenz2} and Lempert and Sz\H{o}ke \cite{LS},
$T^{\ast}\left(  X\right)  $ also has a canonical ``adapted'' complex
structure, defined in a neighborhood of the zero section. This complex
structure is compatible with the symplectic structure and so defines a
K\"{a}hler polarization on an open set in $T^{\ast}\left(  X\right)  .$

This paper considers the special case in which $X$ is a Lie group $K$ with a
bi-invariant Riemannian metric. Lie groups that admit a bi-invariant metric
are said to be of ``compact type''; these are precisely the groups of the form
$\left(  \text{compact}\right)  \times\mathbb{R}^{n}.$ In this special case,
the adapted complex structure is defined on all of $T^{\ast}(K),$ so $T^{\ast
}(K)$ has two polarizations, the vertical polarization and the K\"{a}hler
polarization coming from the adapted complex structure. If $K=\mathbb{R}^{n}$
then the complex structure is just the usual one on $T^{\ast}\left(
\mathbb{R}^{n}\right)  =\mathbb{R}^{2n}=\mathbb{C}^{n}.$

There are two main results, generalizing what is known in the $\mathbb{R}^{n}
$ case. First, the K\"{a}hler-polarized Hilbert space constructed over
$T^{\ast}(K)$ is naturally identifiable with the generalized Segal--Bargmann
space defined in \cite{H1} in terms of heat kernels. Second, the pairing map
between the vertically polarized and the K\"{a}hler-polarized Hilbert space
over $T^{\ast}(K)$ coincides (up to a constant) with the generalized
Segal--Bargmann transform of \cite{H1}. Thus by \cite[Thm. 2]{H1} a constant
multiple of the pairing map is unitary in this case. Both of these results
hold only if one includes the ``half-form correction'' in the construction of
the K\"{a}hler-polarized Hilbert space. In the case $K=\mathbb{R}^{n}$
everything reduces to the ordinary Segal--Bargmann space and the
Segal--Bargmann transform (Section 5).

The results are surprising for two reasons. First, the constructions in
\cite{H1} involve heat kernels, whereas geometric quantization seems to have
nothing to do with heat kernels or the heat equation. Second, in the absence
of something like the Stone--von Neumann theorem there does not seem to be any
reason that pairing maps \textit{ought} to be unitary. The discussion in
Section 4 gives some partial explanation for the occurrence of the heat
kernel. (See also \cite{JL}.)

If one considers Yang--Mills theory over a space-time cylinder, in the
temporal gauge, the ``unreduced phase space'' is a certain
infinite-dimensional linear space of connections. The reduced phase space,
obtained by ``reducing'' by a suitable gauge group, is the finite-dimensional
symplectic manifold $T^{\ast}(K),$ where $K$ is the structure group for the
Yang-Mills theory. Thus the symplectic manifold $T^{\ast}(K)$ considered here
can also be viewed as the ``symplectic quotient'' of an infinite-dimensional
linear space by an infinite-dimensional group. It is reasonable to ask whether
``quantization commutes with reduction,'' that is, whether one gets the same
results by first quantizing and then reducing as by first reducing and then
quantizing. Surprisingly (to me), the answer in this case is yes, as described
in Section 3.

I conclude this introduction by discussing two additional points. First, it is
reasonable to consider the more general situation where the group $K$ is
allowed to be a symmetric space of compact type. In that case the geometric
quantization constructions make perfect sense, but the main results of this
paper do not hold. Specifically, the K\"{a}hler-polarized Hilbert space does
not coincide with the heat kernel Hilbert space of M. Stenzel \cite{St}, and
I\ do not know whether the pairing map of geometric quantization is unitary.
This discrepancy reflects special properties that compact Lie groups have
among all compact symmetric spaces. See the discussion at the end of Section 2.3.

Second, one could attempt to construct a momentum Hilbert space for $T^{\ast
}(K).$ In the case $K=\mathbb{R}^{n}$ this may be done by considering the
natural horizontal polarization. The pairing map between the vertically
polarized and horizontally polarized Hilbert spaces is in this case just the
Fourier transform. By contrast, if $K$ is non-commutative, then there is no
natural horizontal polarization. (For example, the foliation of $T^{\ast}(K) $
into the left orbits of $K$ is not Lagrangian.) Thus, even though there is a
sort of momentum representation given by the Peter--Weyl theorem, it does not
seem possible to obtain a momentum representation by means of geometric quantization.

It is a pleasure to thank Bruce Driver for valuable discussions, Dan Freed for
making an important suggestion regarding to the half-form correction, and
Steve Sontz for making corrections to the manuscript.

\section{The main results}

\subsection{Preliminaries}

Let $K$ be a connected Lie group of \textbf{compact type}. A Lie group is said
to be of compact type if it is locally isomorphic to some compact Lie group.
Equivalently, a Lie group $K$ is of compact type if there exists an inner
product on the Lie algebra of $K$ that is invariant under the adjoint action
of $K$. So $\mathbb{R}^{n}$ is of compact type, being locally isomorphic to a
$d$-torus, and every compact Lie group is of compact type. It can be shown
that every connected Lie group of compact type is isomorphic to a product of
$\mathbb{R}^{n}$ and a connected compact Lie group. So all of the
constructions described here for Lie groups of compact type include as a
special case the constructions for $\mathbb{R}^{n}.$ On the other hand, all
the new information (beyond the $\mathbb{R}^{n}$ case) is contained in the
compact case. See \cite[Chap. II, Sect. 6]{He} (including Proposition 6.8) for
information on Lie groups of compact type.

Let $\frak{k}$ denote the Lie algebra of $K.$ We fix once and for all an inner
product $\left\langle \cdot,\cdot\right\rangle $ on $\frak{k}$ that is
invariant under the adjoint action of $K.$ For example we may take
$K=\mathrm{SU}(n),$ in which case $\frak{k}=\mathrm{su}(n)$ is the space of
skew matrices with trace zero. An invariant inner product on $\frak{k}$ is
$\left\langle X,Y\right\rangle =\operatorname{Re}\left[  \mathrm{trace}\left(
X^{\ast}Y\right)  \right]  .$

Now let $K_{\mathbb{C}}$ be the \textit{complexification} of $K$. If $K$ is
simply connected then the complexification of $K$ is the unique simply
connected Lie group whose Lie algebra $\frak{k}_{\mathbb{C}}$ is
$\frak{k}+i\frak{k}.$ In general, $K_{\mathbb{C}}$ is defined by the following
three properties. First, $K_{\mathbb{C}}$ should be a connected complex Lie
group whose Lie algebra $\frak{k}_{\mathbb{C}}$ is equal to $\frak{k}%
+i\frak{k}.$ Second, $K_{\mathbb{C}}$ should contain $K$ as a closed subgroup
(whose Lie algebra is $\frak{k}\subset\frak{k}_{\mathbb{C}}$). Third, every
homomorphism of $K$ into a complex Lie group $H$ should extend to a
holomorphic homomorphism of $K_{\mathbb{C}}$ into $H.$ The complexification of
a connected Lie group of compact type always exists and is unique. (See
\cite[Sect. 3]{H1}.)

\begin{example}
If $K=\mathbb{R}^{n}$ then $K_{\mathbb{C}}=\mathbb{C}^{n}.$ If $K=\mathrm{SU}%
(n)$ then $K_{\mathbb{C}}=\mathrm{SL}(n;\mathbb{C}).$ If $K=\mathrm{SO}(n)$
then $K_{\mathbb{C}}=\mathrm{SO}(n;\mathbb{C}).$ In the first two examples,
$K$ and $K_{\mathbb{C}}$ are simply connected. In the last example, neither $K
$ nor $K_{\mathbb{C}}$ is simply connected.
\end{example}

We have the following structure result for Lie groups of compact type. This
result is a modest strengthening of Corollary 2.2 of \cite{Dr} and allows all
the relevant results for Lie groups of compact type to be reduced to two
cases, the compact case and the $\mathbb{R}^{n}$ case.

\begin{proposition}
\label{structure.prop}Suppose that $K$ is a connected Lie group of compact
type, with a fixed Ad-invariant inner product on its Lie algebra $\frak{k}.$
Then there exists a isomorphism $K\cong H\times\mathbb{R}^{n},$ where $H$ is
compact and where the associated Lie algebra isomorphism $\frak{k}%
=\frak{h}+\mathbb{R}^{n}$ is orthogonal.
\end{proposition}

The proof of this result is given in an appendix.

\subsection{Prequantization}

We let $\theta$ be the canonical 1-form on $T^{\ast}(K),$ normalized so that
in the usual sort of coordinates we have
\[
\theta=\sum p_{k}\,dq_{k}.
\]
We then let $\omega$ be the canonical 2-form on $T^{\ast}(K),$ which I
normalize as $\omega=-d\theta,$ so that in coordinates $\omega=\Sigma
dq_{k}\wedge dp_{k}.$ We then consider a trivial complex line bundle $L$ on
$T^{\ast}(K)$%
\[
L=T^{\ast}(K)\times\mathbb{C}%
\]
with trivial Hermitian structure. Sections of this bundle are thus just
functions on $T^{\ast}(K).$ We define a connection (or covariant derivative)
on $L$ by
\begin{equation}
\nabla_{X}=X-\frac{1}{i\hbar}\theta\left(  X\right)  .\label{covariant}%
\end{equation}
Note that the connection, and hence all subsequent constructions, depends on
$\hbar$ (Planck's constant). The curvature of this connection is given by
\[
\left[  \nabla_{X},\nabla_{Y}\right]  -\nabla_{\left[  X,Y\right]  }=\frac
{1}{i\hbar}\omega\left(  X,Y\right)  .
\]

We let $\varepsilon$ denote the Liouville volume form on $T^{\ast}(K),$ given
by
\[
\varepsilon=\frac{1}{n!}\omega^{n},
\]
where $n=\dim K=\left(  1/2\right)  \dim T^{\ast}(K).$ Integrating this form
gives the associated Liouville volume measure. Concretely we have the
identification
\begin{equation}
T^{\ast}(K)\cong K\times\frak{k}\label{tkform}%
\end{equation}
by means of left-translation and the inner product on $\frak{k}.$ Under this
identification we have \cite[Lem. 4]{H3}
\begin{equation}
\int_{T^{\ast}(K)}f\,\varepsilon=\int_{\frak{k}}\int_{K}f\left(  x,Y\right)
\,dx\,dY\label{epsilonform}%
\end{equation}
where $dx$ is Haar measure on $K,$ normalized to coincide with the Riemannian
volume measure, and $dY$ is Lebesgue measure on $\frak{k},$ normalized by
means of the inner product. The prequantum Hilbert space is then the space of
sections of $L$ that are square integrable with respect to $\varepsilon.$ This
space may be identified with $L^{2}\left(  T^{\ast}(K),\varepsilon\right)  .$

One motivation for this construction is the existence of a natural mapping
$\mathcal{Q}$ from functions on $T^{\ast}(K)$ into the space of symmetric
operators on the prequantum Hilbert space, satisfying $[\mathcal{Q}\left(
f\right)  ,\mathcal{Q}\left(  g\right)  ]=-i\hbar\mathcal{Q}\left(  \left\{
f,g\right\}  \right)  ,$ where $\left\{  f,g\right\}  $ is the Poisson
bracket. Explicitly, $\mathcal{Q}\left(  f\right)  =i\hbar\nabla_{X_{f}}+f,$
where $X_{f}$ is the Hamiltonian vector field associated to $f.$ This
``prequantization map'' will not play an important role in this paper. See
\cite[Chap. 8]{Wo} for more information.

\subsection{The K\"{a}hler-polarized subspace}

Let me summarize what the results of this subsection will be. The cotangent
bundle $T^{\ast}(K)$ has a natural complex structure that comes by identifying
it with the ``complexification'' of $K.$ This complex structure allows us to
define a notion of K\"{a}hler-polarized sections of the bundle $L.$ There
exists a natural trivializing polarized section $s_{0}$ such that every other
polarized section is a holomorphic function times $s_{0}.$ The K\"{a}%
hler-polarized Hilbert space is then identifiable with an $L^{2}$ space of
holomorphic functions on $T^{\ast}(K),$ where the measure is the Liouville
measure times $\left|  s_{0}\right|  ^{2}.$ We then consider the ``half-form''
bundle $\delta_{1}.$ The half-form corrected K\"{a}hler Hilbert space is the
space of polarized sections of $L\otimes\delta_{1}.$ This may be identified
with an $L^{2}$ space of holomorphic functions on $T^{\ast}(K),$ where now the
measure is the Liouville measure times $\left|  s_{0}\right|  ^{2}\left|
\beta_{0}\right|  ^{2},$ where $\beta_{0}$ is a trivializing polarized section
of $\delta_{1}.$ The main result is that this last measure coincides up to a
constant with the $K$-invariant heat kernel measure on $T^{\ast}(K)$
introduced in \cite{H1}. Thus the half-form-corrected K\"{a}hler-polarized
Hilbert space of geometric quantization coincides (up to a constant) with the
generalized Segal--Bargmann space of \cite[Thm. 2]{H1}.

We let $K_{\mathbb{C}}$ denote the complexification of $K,$ as described in
Section 2.1, and we let $T^{\ast}(K)$ denote the cotangent bundle of $K.$
There is a diffeomorphism of $T^{\ast}(K)$ with $K_{\mathbb{C}}$ as follows.
We identify $T^{\ast}(K)$ with $K\times\frak{k}^{\ast}$ by means of
left-translation and then with $K\times\frak{k}$ by means of the inner product
on $\frak{k}.$ We consider the map $\Phi:K\times\frak{k}\rightarrow
K_{\mathbb{C}}$ given by
\begin{equation}
\Phi\left(  x,Y\right)  =xe^{iY},\quad x\in K,\,\,Y\in\frak{k.}\label{phimap}%
\end{equation}
The map $\Phi$ is a diffeomorphism. If we use $\Phi$ to transport the complex
structure of $K_{\mathbb{C}}$ to $T^{\ast}(K),$ then the resulting complex
structure on $T^{\ast}(K)$ is compatible with the symplectic structure on
$T^{\ast}(K),$ so that $T^{\ast}(K)$ becomes a K\"{a}hler manifold. (See
\cite[Sect. 3]{H3}.)

Consider the function $\kappa:T^{\ast}(K)\rightarrow\mathbb{R}$ given by
\begin{equation}
\kappa\left(  x,Y\right)  =\left|  Y\right|  ^{2}.\label{kappa}%
\end{equation}
This function is a \textit{K\"{a}hler potential} for the complex structure on
$T^{\ast}(K)$ described in the previous paragraph. Specifically we have
\begin{equation}
\operatorname{Im}\left(  \bar{\partial}\kappa\right)  =\theta.\label{kappa1}%
\end{equation}
Then because $\omega=-d\theta$ it follows that
\begin{equation}
i\partial\bar{\partial}\kappa=\omega.\label{kappa2}%
\end{equation}
An important feature of this situation is the natural explicit form of the
K\"{a}hler potential. This formula for $\kappa$ comes as a special case of the
general construction of Guillemin--Stenzel \cite[Sect. 5]{GStenz1} and
Lempert--Sz\H{o}ke \cite[Cor. 5.5]{LS}. It this case one can compute directly
that $\kappa$ satisfies (\ref{kappa1}) and (\ref{kappa2}) (see the first appendix).

We define a smooth section $s$ of $L$ to be \textit{K\"{a}hler-polarized} if
\[
\nabla_{X}s=0
\]
for all vectors of type $\left(  0,1\right)  .$ Equivalently $s$ is polarized
if $\nabla_{\partial/\partial\bar{z}_{k}}s=0$ for all $k,$ in holomorphic
local coordinates. The \textit{K\"{a}hler-polarized Hilbert space} is then the
space of square-integrable K\"{a}hler-polarized sections of $L.$ (See
\cite[Sect. 9.2]{Wo}.)

\begin{proposition}
\label{polarized.prop}If we think of sections $s$ of $L$ as functions on
$T^{\ast}(K)$ then the K\"{a}hler-polarized sections are precisely the
functions $s$ of the form
\[
s=F\,e^{-\left|  Y\right|  ^{2}/2\hbar},
\]
with $F$ holomorphic and $\left|  Y\right|  ^{2}=\kappa\left(  x,Y\right)  $
the K\"{a}hler potential (\ref{kappa}). The notion of holomorphic is via the
identification (\ref{phimap}) of $T^{\ast}(K)$ with $K_{\mathbb{C}}.$
\end{proposition}

\textit{Proof}. If we work in holomorphic local coordinates $z_{1}%
,\cdots,z_{n}$ then we want sections $s$ such that $\nabla_{\partial
/\partial\bar{z}_{k}}s=0$ for all $k.$ The condition (\ref{kappa1}) on
$\kappa$ says that in these coordinates
\[
\theta=\frac{1}{2i}\sum_{k}\left(  \frac{\partial\kappa}{\partial\bar{z}_{k}%
}d\bar{z}_{k}-\frac{\partial\kappa}{\partial z_{k}}dz_{k}\right)  .
\]
So
\[
\theta\left(  \frac{\partial}{\partial\bar{z}_{k}}\right)  =\frac{1}{2i}%
\frac{\partial\kappa}{\partial\bar{z}_{k}}.
\]
Then we get, using the definition (\ref{covariant}) of the covariant
derivative,
\begin{align*}
\nabla_{\partial/\partial\bar{z}_{k}}e^{-\kappa/2\hbar}  & =\frac{\partial
}{\partial\bar{z}_{k}}e^{-\kappa/2\hbar}-\frac{1}{i\hbar}\theta\left(
\frac{\partial}{\partial\bar{z}_{k}}\right)  e^{-\kappa/2\hbar}\\
& =\left(  -\frac{1}{2\hbar}\frac{\partial\kappa}{\partial\bar{z}_{k}}%
-\frac{1}{i\hbar}\frac{1}{2i}\frac{\partial\kappa}{\partial\bar{z}_{k}%
}\right)  e^{-\kappa/2\hbar}=0.
\end{align*}

Now any smooth section $s$ can be written uniquely as $s=F\exp\left(
-\kappa/2\hbar\right)  ,$ where $F$ is a smooth complex-valued function. Such
a section is polarized precisely if
\begin{align*}
0  & =\nabla_{\partial/\partial\bar{z}_{k}}\left(  F\,e^{-\kappa/2\hbar
}\right) \\
& =\frac{\partial F}{\partial\bar{z}_{k}}e^{-\kappa/2\hbar}+F\,\nabla
_{\partial/\partial\bar{z}_{k}}e^{-\kappa/2\hbar}\\
& =\frac{\partial F}{\partial\bar{z}_{k}}e^{-\kappa/2\hbar}%
\end{align*}
for all $k,$ that is, precisely if $F$ is holomorphic.$\,\square$

The norm of a polarized section $s$ (as in Proposition \ref{polarized.prop})
is computed as
\begin{align*}
\left\|  s\right\|  ^{2}  & =\int_{T^{\ast}(K)}\left|  F\right|
^{2}e^{-\kappa/\hbar}\,\varepsilon\\
& =\int_{\frak{k}}\int_{K}\left|  F\left(  xe^{iY}\right)  \right|
^{2}e^{-\left|  Y\right|  ^{2}/\hbar}\,dx\,dY.
\end{align*}
Here $F$ is a holomorphic function on $K_{\mathbb{C}}$ which we are
``transporting'' to $T^{\ast}(K)$ by means of the map $\Phi\left(  x,Y\right)
=xe^{iY}.$ (Recall (\ref{tkform}) and (\ref{epsilonform}).) Thus if we
identify the section $s$ with the holomorphic function $F,$ the K\"{a}%
hler-polarized Hilbert space will be identified with
\[
\mathcal{H}L^{2}(T^{\ast}(K),e^{-\left|  Y\right|  ^{2}/\hbar}\varepsilon)
\]
Here $\varepsilon$ is the Liouville volume measure and $\mathcal{H}L^{2}$
denotes the space of holomorphic functions that are square-integrable with
respect to the indicated measure.

In Section 7 of \cite{H4} I compared the measure $e^{-\left|  Y\right|
^{2}/\hbar}\varepsilon$ to the ``$K$-invariant heat kernel measure''
$\nu_{\hbar}$ on $K_{\mathbb{C}}\cong T^{\ast}(K).$ The measure $\nu_{\hbar}$
is the one that is used in the generalized Segal--Bargmann transform of
\cite[Thm. 2]{H1}. In the commutative case the two measures agree up to a
constant. However, in the non-commutative case the two measures differ by a
non-constant function of $Y,$ and it is easily seen that this discrepancy
cannot be eliminated by choosing a different trivializing polarized section of
$L.$ In the remainder of this section we will see that this discrepancy
between the heat kernel measure and the geometric quantization measure can be
eliminated by the ``half-form correction.'' I am grateful to Dan Freed for
suggesting to me that this could be the case.

We now consider the canonical bundle for $T^{\ast}(K)$ relative to the complex
structure obtained from $K_{\mathbb{C}}.$ The canonical bundle is the complex
line bundle whose sections are complex-valued $n$-forms of type $\left(
n,0\right)  .$ The forms of type $\left(  n,0\right)  $ may be described as
those $n$-forms $\alpha$ for which
\[
X\lrcorner\alpha=0
\]
for all vectors of type $\left(  0,1\right)  .$ We then define the polarized
sections of the canonical bundle to be the $\left(  n,0\right)  $-forms
$\alpha$ such that
\[
X\lrcorner d\alpha=0
\]
for all vector fields of type $\left(  0,1\right)  .$ (Compare \cite[Eq.
(9.3.1)]{Wo}.) These are nothing but the holomorphic $n$-forms. We define a
Hermitian structure on the canonical bundle by defining for an $\left(
n,0\right)  $-form $\alpha$%
\[
\left|  \alpha\right|  ^{2}=\frac{\bar{\alpha}\wedge\alpha}{b\,\varepsilon}.
\]
Here the ratio means the only thing that is reasonable: $\left|
\alpha\right|  ^{2}$ is the unique function such that $\left|  \alpha\right|
^{2}b\varepsilon=\bar{\alpha}\wedge\alpha.$ The constant $b$ should be chosen
in such a way as to make $\left|  \alpha\right|  ^{2}$ positive; we may take
$b=(2i)^{n}(-1)^{n(n-1)/2}.$

In this situation the canonical bundle may be trivialized as follows. We think
of $T^{\ast}(K)$ as $K_{\mathbb{C}},$ since at the moment the symplectic
structure is not relevant. If $Z_{1},\cdots,Z_{n}$ are linearly independent
left-invariant holomorphic 1-forms on $K_{\mathbb{C}}$ then their wedge
product is a nowhere-vanishing holomorphic $n$-form.

We now choose a square root $\delta_{1}$ of the canonical bundle in such a way
that there exists a smooth section of $\delta_{1}$ whose square is
$Z_{1}\wedge\cdots\wedge Z_{n}.$ This section of $\delta_{1}$ will be denoted
by the mnemonic $\sqrt{Z_{1}\wedge\cdots\wedge Z_{n}}.$ There then exists a
unique notion of polarized sections of $\delta_{1}$ such that 1) a locally
defined, smooth, nowhere-zero section $\nu$ of $\delta_{1}$ is polarized if
and only if $\nu^{2}$ is a polarized section of the canonical bundle, and 2)
if $\nu$ is a locally defined, nowhere-zero, polarized section of $\delta_{1}$
and $F$ is a smooth function, then $F\nu$ is polarized if and only if $F$ is
holomorphic. (See \cite[p. 186]{Wo}.) Concretely the polarized sections of
$\delta_{1}$ are of the form
\[
s=F\left(  g\right)  \sqrt{Z_{1}\wedge\cdots\wedge Z_{n}}%
\]
with $F$ a holomorphic function on $K_{\mathbb{C}}.$ The absolute value of
such a section is defined as
\[
\left|  s\right|  ^{2}:=\sqrt{\left(  s^{2},s^{2}\right)  }=\left|  F\right|
^{2}\sqrt{\frac{\bar{Z}_{1}\wedge\cdots\wedge\bar{Z}_{n}\wedge Z_{1}%
\wedge\cdots\wedge Z_{n}}{a\,\varepsilon}}.
\]

Now the \textit{half-form corrected K\"{a}hler-polarized Hilbert space} is the
space of square-integrable polarized sections of $L\otimes\delta_{1}.$ (The
polarized sections of $L\otimes\delta_{1}$ are precisely those that can be
written locally as the product of a polarized section of $L$ and a polarized
section of $\delta_{1}.$) Such sections are precisely those that can be
expressed as
\begin{equation}
s=F\,e^{-\left|  Y\right|  ^{2}/2\hbar}\otimes\sqrt{Z_{1}\wedge\cdots\wedge
Z_{n}}\label{sbform2}%
\end{equation}
with $F$ holomorphic. The norm of such a section is computed as
\[
\left\|  s\right\|  ^{2}=\int_{T^{\ast}(K)}\left|  F\right|  ^{2}e^{-\left|
Y\right|  ^{2}/\hbar}\eta\,\varepsilon,
\]
where $\eta$ is the function given by
\begin{equation}
\eta=\sqrt{\frac{\bar{Z}_{1}\wedge\cdots\wedge\bar{Z}_{n}\wedge Z_{1}%
\wedge\cdots\wedge Z_{n}}{b\,\varepsilon}},\label{half.form}%
\end{equation}
and where $b=(2i)^{n}(-1)^{n(n-1)/2}.$ We may summarize the preceding
discussion in the following theorem.

\begin{theorem}
\label{sbthm2}If we write elements of the half-form corrected K\"{a}hler
Hilbert space in the form (\ref{sbform2}) then this Hilbert space may be
identified with
\[
\mathcal{H}L^{2}(T^{\ast}(K),\gamma_{\hbar})
\]
where $\gamma_{\hbar}$ is the measure given by
\[
\gamma_{\hbar}=e^{-\left|  Y\right|  ^{2}/\hbar}\eta\,\varepsilon.
\]
Here $\varepsilon$ is the canonical volume form on $T^{\ast}(K),$ $\left|
Y\right|  ^{2}$ is the K\"{a}hler potential (\ref{kappa}), and $\eta$ is the
``half-form correction'' defined in (\ref{half.form}) and given explicitly in
(\ref{eta.form}) below. Here as elsewhere $\mathcal{H}L^{2}$ denotes the space
of square-integrable holomorphic functions.
\end{theorem}

Note that $\bar{Z}_{1}\wedge\cdots\wedge\bar{Z}_{n}\wedge Z_{1}\wedge
\cdots\wedge Z_{n}$ is a left-invariant $2n$-form on $K_{\mathbb{C}},$ so that
the associated measure is simply a multiple of Haar measure on $K_{\mathbb{C}%
}.$ Meanwhile $\varepsilon$ is just the Liouville volume form on $T^{\ast
}(K).$ Thus $\eta$ is the square root of the density of Haar measure with
respect to Liouville measure, under our identification of $K_{\mathbb{C}}$
with $T^{\ast}(K)$. Both measures are $K$-invariant, so in our $\left(
x,Y\right)  $ coordinates on $T^{\ast}(K),$ $\eta$ will be a function of $Y$
only. By \cite[Lem. 5]{H3} we have that $\eta\left(  Y\right)  $ is the unique
Ad-$K$-invariant function on $\frak{k}$ such that in a maximal abelian
subalgebra
\begin{equation}
\eta\left(  Y\right)  =\prod_{\alpha\in R^{+}}\frac{\sinh\alpha\left(
Y\right)  }{\alpha\left(  Y\right)  },\label{eta.form}%
\end{equation}
where $R^{+}$ is a set of positive roots.

Meanwhile there is the ``$K$-invariant heat kernel measure'' $\nu_{\hbar}$ on
$K_{\mathbb{C}}\cong T^{\ast}(K),$ used in the construction of the generalized
Segal--Bargmann transform in \cite[Thm. 2]{H1}. When written in terms of the
polar decomposition $g=xe^{iY},$ $\nu_{\hbar}$ is given explicitly by
\[
d\nu_{\hbar}=\left(  \pi\hbar\right)  ^{-n/2}e^{-\left|  \rho\right|
^{2}\hbar}e^{-\left|  Y\right|  ^{2}/\hbar}\eta\left(  Y\right)  \,dx\,dY.
\]
(See \cite[Eq. (13)]{H3}.) Here $\rho$ is half the sum of the positive roots
for the group $K.$ Thus apart from an overall constant, the measure $T^{\ast
}(K)$ coming from geometric quantization coincides exactly with the heat
kernel measure of \cite{H1}. So we have proved the following result.

\begin{theorem}
For each $\hbar>0$ there exists a constant $c_{\hbar}$ such that the measure
$\gamma_{\hbar}$ coming from geometric quantization and the heat kernel
measure $\nu_{\hbar}$ are related by
\[
\nu_{\hbar}=c_{\hbar}\gamma_{\hbar}%
\]
where
\[
c_{\hbar}=\left(  \pi\hbar\right)  ^{-n/2}e^{-\left|  \rho\right|  ^{2}\hbar}%
\]
and where $\rho$ is half the sum of the positive roots for the group $K.$
\end{theorem}

Let us try to understand, at least in part, the seemingly miraculous agreement
between these two measures. (See also Section 4.) The cotangent bundle
$T^{\ast}(K)$ has a complex structure obtained by identification with
$K_{\mathbb{C}}.$ The metric tensor on $K$ then has an analytic continuation
to a holomorphic $n$-tensor on $T^{\ast}(K).$ The restriction of the
analytically continued metric tensor to the fibers of $T^{\ast}(K)$ is the
negative of a Riemannian metric $g.$ Each fiber, with this metric, is
isometric to the non-compact symmetric space $K_{\mathbb{C}}/K.$ (See
\cite{St}.) This reflects the well-known duality between compact and
non-compact symmetric spaces. Each fiber is also identified with $\frak{k},$
and under this identification the Riemannian volume measure with respect to
$g$ is given by
\[
\sqrt{g}dY=\eta\left(  Y\right)  ^{2}\,dY.
\]
That is, the ``half-form factor'' $\eta$ is simply the square root of the
Jacobian of the exponential mapping for $K_{\mathbb{C}}/K.$

Now on any Riemannian manifold the heat kernel measure (at a fixed base point,
written in exponential coordinates) has an asymptotic expansion of the form
\begin{equation}
d\mu_{\hbar}\left(  Y\right)  \sim\left(  \pi\hbar\right)  ^{-n/2}e^{-\left|
Y\right|  ^{2}/\hbar}\left(  j^{1/2}\left(  Y\right)  +ta_{1}\left(  Y\right)
+t^{2}a_{2}\left(  Y\right)  +\cdots\right)  \,dY.\label{mp}%
\end{equation}
Here $j\left(  Y\right)  $ is the Jacobian of the exponential mapping, also
known as the Van Vleck--Morette determinant. (I have written $\hbar$ for the
time variable and normalized the heat equation to be $du/dt=(1/4)\Delta u.$)
Note that this is the expansion for the heat kernel \textit{measure}; in the
expansion of the heat kernel \textit{function} one has $j^{-1/2}$ instead of
$j^{1/2}.$

In the case of the manifold $K_{\mathbb{C}}/K$ we have a great simplification.
All the higher terms in the series are just constant multiples of $j^{1/2}$
and we get an exact convergent expression of the form
\begin{equation}
d\mu_{\hbar}\left(  Y\right)  =\left(  \pi\hbar\right)  ^{-n/2}e^{-\left|
Y\right|  ^{2}/\hbar}j^{1/2}\left(  Y\right)  f\left(  t\right)
\,dY.\label{mp2}%
\end{equation}
Here explicitly $f\left(  t\right)  =\exp(-\left|  \rho\right|  ^{2}t),$ where
$\rho$ is half the sum of the positive roots. The measure $\nu_{\hbar}$ in
\cite{H1} is then simply this measure times the Haar measure $dx$ in the
$K$-directions. So we have
\[
d\nu_{\hbar}=e^{-\left|  \rho\right|  ^{2}t}\left(  \pi\hbar\right)
^{-n/2}e^{-\left|  Y\right|  ^{2}/\hbar}j^{1/2}\left(  Y\right)  \,dx\,dY.
\]

So how does geometric quantization produce a multiple of $\nu_{\hbar}$? The
Gaussian factor in $\nu_{\hbar}$ comes from the simple explicit form of the
K\"{a}hler potential. The factor of $j^{1/2}$ in $\nu_{\hbar}$ is the
half-form correction--that is, $j^{1/2}(Y)=\eta(Y)$

If we begin with a general compact symmetric space $X$ then much of the
analysis goes through: $T^{\ast}(X)$ has a natural complex structure, $\left|
Y\right|  ^{2}$ is a K\"{a}hler potential, and the fibers are identifiable
with non-compact symmetric spaces. (See \cite[p. 48]{St}.) Furthermore, the
half-form correction is still the square root of the Jacobian of the
exponential mapping. What goes wrong is that the heat kernel expansion
(\ref{mp}) does not simplify to an expression of the form (\ref{mp2}). So the
heat kernel measure used in \cite{St} and the measure coming from geometric
quantization will not agree up to a constant. Nevertheless the two measures do
agree ``to leading order in $\hbar$.''

I do not know whether the geometric quantization pairing map is unitary in the
case of general compact symmetric spaces $X.$ There is, however, a unitary
Segal--Bargmann-type transform, given in terms of heat kernels and described
in \cite{St}.

\subsection{The vertically polarized Hilbert space}

After much sound and fury, the vertically polarized Hilbert space will be
identified simply with $L^{2}\left(  K,dx\right)  $, where $dx$ is Haar
measure on $K.$ Nevertheless, the fancy constructions described below are
important for two reasons. First, the vertically polarized Hilbert space does
not depend on a choice of measure on $K.$ The Hilbert space is really a space
of ``half-forms.'' If one chooses a smooth measure $\mu$ on $K$ (with nowhere
vanishing density with respect to Lebesgue measure in each local coordinate
system) then this choice gives an identification of the vertically polarized
Hilbert space with $L^{2}(K,\mu).$ Although Haar measure is the obvious choice
for $\mu,$ the choice of measure is needed only to give a concrete realization
of the space as an $L^{2}$ space; the vertically polarized Hilbert space
exists independently of this choice. Second, the description of the vertically
polarized Hilbert space as space of half-forms will be essential to the
construction of the pairing map in Section 2.5.

The following description follows Section 9.3 of \cite{Wo}. Roughly speaking
our Hilbert space will consist of objects whose squares are $n$-forms on
$T^{\ast}(K)$ that are constant along the fibers and thus descend to $n$-forms
on $K.$ The norm of such an object is computed by squaring and then
integrating the resulting $n$-form over $K.$

We consider sections of $L$ that are covariantly constant in the directions
parallel to the fibers of $T^{\ast}(K).$ Note that each fiber of $T^{\ast}(K)$
is a Lagrangian submanifold of $T^{\ast}(K),$ so that $T^{\ast}(K)$ is
naturally foliated into Lagrangian submanifolds. Suppose that $X$ is a tangent
vector to $T^{\ast}(K)$ that is parallel to one of the fibers. Then it is
easily seen that $\theta\left(  X\right)  =0,$ where $\theta$ is the canonical
1-form on $T^{\ast}(K).$ Thus, recalling the definition (\ref{covariant}) of
the covariant derivative and thinking of the sections of $L$ as functions on
$T^{\ast}(K),$ the vertically polarized sections are simply the functions that
are constant along the fibers. Such a section cannot be square-integrable with
respect to the Liouville measure (unless it is zero almost everywhere). This
means that we cannot construct the vertically polarized Hilbert space as a
subspace of the prequantum Hilbert space.

We consider, then, the canonical bundle of $T^{\ast}(K)$ relative to the
vertical polarization. This is the \textit{real} line bundle whose sections
are $n$-forms $\alpha$ such that
\begin{equation}
X\lrcorner\alpha=0\label{canonical4}%
\end{equation}
for all vectors parallel to the fibers of $T^{\ast}(K).$ We call such a
section polarized if in addition we have
\begin{equation}
X\lrcorner d\alpha=0\label{canpolarized4}%
\end{equation}
for all vectors $X$ parallel to the fibers. (See \cite[Eq. (9.3.1)]{Wo}.)

Now let $Q$ be the space of fibers (or the space of leaves of our Lagrangian
foliation). Clearly $Q$ may be identified with $K$ itself, the ``configuration
space'' corresponding to the ``phase space'' $T^{\ast}(K)$. Let $pr:T^{\ast
}(K)\rightarrow K$ be the projection map. It is not hard to verify that if
$\alpha$ is a $n$-form on $T^{\ast}(K)$ satisfying (\ref{canonical4}) and
(\ref{canpolarized4}) then there exists a unique $n$-form $\beta$ on $K$ such
that
\[
\alpha=pr^{\ast}\left(  \beta\right)  .
\]
We may think of such an $n$-form $\alpha$ as being constant along the fibers,
so that it descends unambiguously to an $n$-form $\beta$ on $K.$ In this way
the polarized sections of the canonical bundle may be identified with
$n$-forms on $K.$

Since $K$ is a Lie group it is orientable. So let us pick an orientation on
$K,$ which we think of as an equivalence class of nowhere-vanishing $n$-forms
on $K.$ Then if $\beta$ is a nowhere-vanishing oriented $n$-form on $K,$ we
define the ``positive'' part of each fiber of the canonical bundle to be the
half-line in which $pr^{\ast}\left(  \beta\right)  $ lies. We may then
construct a unique trivial real line bundle $\delta_{2}$ such that 1) the
square of $\delta_{2}$ is the canonical bundle and 2) if $\gamma$ is a
nowhere-vanishing section of $\delta_{2}$ then $\gamma^{2}$ lies in the
positive part of the canonical bundle. We have a natural notion of polarized
sections of $\delta_{2},$ such that 1) a locally defined, smooth, nowhere-zero
section $\nu$ of $\delta_{2}$ is polarized if and only if $\nu^{2}$ is a
polarized section of the canonical bundle and 2) if $\nu$ is a locally
defined, nowhere-zero, polarized section of $\delta_{2}$ and $f$ is a smooth
function, then $f\nu$ is polarized if and only if $f$ is constant along the fibers.

Now let $\beta$ be any nowhere vanishing oriented $n$-form on $K.$ Then there
exists a polarized section of $\delta_{2}$ (unique up to an overall sign)
whose square is $pr^{\ast}\left(  \beta\right)  .$ This section is denoted
$\sqrt{pr^{\ast}\left(  \beta\right)  }.$ Any other polarized section of
$\delta_{2}$ is then of the form
\[
f\left(  x\right)  \sqrt{pr^{\ast}\left(  \beta\right)  },
\]
where $f\left(  x\right)  $ denotes a real-valued function on $T^{\ast}(K)$
that is constant along the fibers.

Finally we consider polarized sections of $L\otimes\delta_{2},$ i.e. those
that are locally the product of a vertically polarized section of $L$ and a
polarized section of $\delta_{2}.$ These are precisely the sections that can
be expressed in the form
\[
s=f\left(  x\right)  \otimes\sqrt{pr^{\ast}\left(  \beta\right)  },
\]
where $f$ is a complex-valued function on $T^{\ast}(K)$ that is constant along
the fibers. The norm of such a section is computed as
\[
\left\|  s\right\|  ^{2}=\int_{K}\left|  f\left(  x\right)  \right|  ^{2}%
\beta.
\]
It is easily seen that this expression for $\left\|  s\right\|  $ is
independent of the choice of $\beta.$ Note that the integration is over the
quotient space $K,$ not over $T^{\ast}(K).$

In particular we may choose linearly independent left-invariant 1-forms
$\eta_{1},\cdots,\eta_{n}$ on $K$ in such a way that $\eta_{1}\wedge
\cdots\wedge\eta_{n}$ is oriented. Then every polarized section of
$L\otimes\delta_{2}$ is of the form
\[
s=f\left(  x\right)  \otimes\sqrt{pr^{\ast}\left(  \eta_{1}\wedge\cdots
\wedge\eta_{n}\right)  }%
\]
and the norm of a section is computable as
\begin{align}
\left\|  s\right\|  ^{2}  & =\int_{K}\left|  f\left(  x\right)  \right|
^{2}\eta_{1}\wedge\cdots\wedge\eta_{n}\nonumber\\
& =\int_{K}\left|  f\left(  x\right)  \right|  ^{2}\,dx,\label{vert.l2}%
\end{align}
where $dx$ is Haar measure on $K.$ Thus we may identify the vertically
polarized Hilbert space with $L^{2}(K,dx).$ More precisely, if we assume up to
now that all sections are smooth, then we have the subspace of $L^{2}\left(
K,dx\right)  $ consisting of smooth functions. The vertically polarized
Hilbert space is then the completion of this space, which is just
$L^{2}\left(  K,dx\right)  .$

\subsection{Pairing}

Geometric quantization gives a way to define a \textit{pairing} between the
K\"{a}hler-polarized and vertically polarized Hilbert spaces, that is, a
sesquilinear map from $H_{K\ddot{a}hler}\times H_{Vertical}$ into
$\mathbb{C}.$ This pairing then induces a linear map between the two spaces,
called the \textit{pairing map}. The main results are: 1) the pairing map
coincides up to a constant with the generalized Segal--Bargmann transform of
\cite{H1}, and 2) a constant multiple of the pairing map is unitary from the
vertically polarized Hilbert space onto the K\"{a}hler-polarized Hilbert space.

Now the elements of the K\"{a}hler-polarized Hilbert space are polarized
sections of $L\otimes\delta_{1}$ and the elements of the vertically polarized
Hilbert space are polarized sections of $L\otimes\delta_{2}.$ Here $\delta
_{1}$ and $\delta_{2}$ are square roots of the canonical bundle for the
K\"{a}hler polarization and the vertical polarization, respectively. The
pairing of the Hilbert spaces will be achieved by appropriately pairing the
sections at each point and then integrating over $T^{\ast}(K)$ with respect to
the canonical volume form $\varepsilon.$ (See \cite[p. 234]{Wo}.)

A polarized section $s_{1}$ of $L\otimes\delta_{1}$ can be expressed as
$s_{1}=f_{1}\otimes\beta_{1},$ where $f_{1}$ is a K\"{a}hler-polarized section
of $L$ and $\beta_{1}$ is a polarized section of $\delta_{1}.$ Similarly, a
polarized section of $L\otimes\delta_{2}$ is expressible as $s_{2}%
=f_{2}\otimes\beta_{2}$ with $f_{2}$ a vertically polarized section of $L$ and
$\beta_{2}$ a polarized section of $\delta_{2}.$ We define a pairing between
$\beta_{1}$ and $\beta_{2}$ by
\[
\left(  \beta_{1},\beta_{2}\right)  =\sqrt{\frac{\overline{\beta_{1}^{2}%
}\wedge\beta_{2}^{2}}{c\,\varepsilon}},
\]
where $c$ is constant which I will take to be $c=(-i)^{n}(-1)^{n(n+1)/2}.$
(This constant is chosen so that things come out nicely in the $\mathbb{R}%
^{n}$ case. See Section 5.) Note that $\beta_{1}^{2}$ and $\beta_{2}^{2}$ are
$n$-forms on $T^{\ast}(K),$ so that $\overline{\beta_{1}^{2}}\wedge\beta
_{2}^{2}$ is a $2n$-form on $T^{\ast}(K).$ Note that $\left(  \beta_{1}%
,\beta_{2}\right)  $ is a complex-valued function on $T^{\ast}(K).$ There are
at most two continuous ways of choosing the sign of the square root, which
differ just by a single overall sign. That there is at least one such choice
will be evident below.

We then define the pairing of two sections $s_{1}$ and $s_{2}$ (as in the
previous paragraph) by
\begin{equation}
\left\langle s_{1},s_{2}\right\rangle _{pair}=\int_{T^{\ast}(K)}\left(
f_{1},f_{2}\right)  \left(  \beta_{1},\beta_{2}\right)  \,\varepsilon
\label{grppair1}%
\end{equation}
whenever the integral is well-defined. Here as usual $\varepsilon$ is the
Liouville volume form on $T^{\ast}(K).$ It is easily seen that this expression
is independent of the decomposition of $s_{i}$ as $f_{i}\otimes\beta_{i}.$ The
quantity $\left(  f_{1},f_{2}\right)  $ is computed using the (trivial)
Hermitian structure on the line bundle $L.$ Although the integral in
(\ref{grppair1}) may not be absolutely convergent in general, there are dense
subspaces of the two Hilbert spaces for which it is. Furthermore, Theorem
\ref{pair.thm} below will show that the pairing can be extended by continuity
to all $s_{1,}s_{2}$ in their respective Hilbert spaces.

Now, we have expressed the polarized sections of $L\otimes\delta_{1}$ in the
form
\[
F\,e^{-\left|  Y\right|  ^{2}/2\hbar}\otimes\sqrt{Z_{1}\wedge\cdots\wedge
Z_{n}}%
\]
where $F$ is a holomorphic function on $K_{\mathbb{C}}$ and $Z_{1}%
,\cdots,Z_{n}$ are left-invariant holomorphic 1-forms on $K_{\mathbb{C}}.$ As
always we identify $K_{\mathbb{C}}$ with $T^{\ast}(K)$ as in (\ref{phimap}).
The function $\left|  Y\right|  ^{2}$ is the K\"{a}hler potential
(\ref{kappa}). We have expressed the polarized section of $L\otimes\delta_{2}$
in the form
\[
f\left(  x\right)  \otimes\sqrt{pr^{\ast}\left(  \eta_{1}\wedge\cdots
\wedge\eta_{n}\right)  },
\]
where $f\left(  x\right)  $ is a function on $T^{\ast}(K)$ that is constant
along the fibers, $\eta_{1},\cdots,\eta_{n}$ are left-invariant 1-forms on
$K,$ and $pr:T^{\ast}(K)\rightarrow K$ is the projection map.

Thus we have the following expression for the pairing :
\begin{equation}
\left\langle F,f\right\rangle _{pair}=\int_{K}\int_{\frak{k}}\overline
{F\left(  xe^{iY}\right)  }f\left(  x\right)  e^{-\left|  Y\right|
^{2}/2\hbar}\zeta\left(  Y\right)  \,dx\,dY,\label{grppair2}%
\end{equation}
where $\zeta$ is the function on $T^{\ast}(K)$ given by
\begin{equation}
\zeta=\sqrt{\frac{\bar{Z}_{1}\wedge\cdots\wedge\bar{Z}_{n}\wedge pr^{\ast
}\left(  \eta_{1}\wedge\cdots\wedge\eta_{n}\right)  }{c\,\varepsilon}%
},\label{zeta.def}%
\end{equation}
where $c=(-i)^{n}(-1)^{n(n+1)/2}.$ I have expressed things in terms of the
functions $F$ and $f,$ and I\ have used the identification (\ref{tkform}) of
$T^{\ast}(K)$ with $K\times\frak{k}.$ It is easily seen that $\zeta\left(
x,Y\right)  $ is independent of $x,$ and so I have written $\zeta\left(
Y\right)  .$

\begin{theorem}
\label{pair.thm}Let us identify the vertically polarized Hilbert space with
$L^{2}\left(  K\right)  $ as in (\ref{vert.l2}) and the K\"{a}hler-polarized
Hilbert space with $\mathcal{H}L^{2}(T^{\ast}(K),\gamma_{\hbar})$ as in
Theorem \ref{sbthm2}. Then there exists a unique bounded linear operator
$\Pi_{\hbar}:L^{2}(K)\rightarrow\mathcal{H}L^{2}(T^{\ast}(K),\gamma_{\hbar})$
such that
\[
\left\langle F,f\right\rangle _{pair}=\left\langle F,\Pi_{\hbar}f\right\rangle
_{\mathcal{H}L^{2}(T^{\ast}(K),\gamma_{\hbar})}=\left\langle \Pi_{\hbar}%
^{\ast}F,f\right\rangle _{L^{2}(K)}%
\]
for all $f\in L^{2}(K)$ and all $F\in\mathcal{H}L^{2}(T^{\ast}(K),\gamma
_{\hbar}).$ We call $\Pi_{\hbar}$ the \textbf{pairing map}.

The pairing map has the following properties.

1) There exists a constant $a_{\hbar}$ such that for any $f\in L^{2}\left(
K\right)  ,$ $\Pi_{\hbar}f$ is the unique holomorphic function on $T^{\ast
}(K)$ whose restriction to $K$ is given by
\[
\left.  \left(  \Pi_{\hbar}f\right)  \right|  _{K}=a_{\hbar}e^{\hbar\Delta
_{K}/2}f.
\]
Equivalently,
\[
\Pi_{\hbar}f\left(  g\right)  =a_{\hbar}\int_{K}\rho_{\hbar}(gx^{-1})f\left(
x\right)  \,dx,\quad g\in K_{\mathbb{C}},
\]
where $\rho_{\hbar}$ is the heat kernel on $K,$ analytically continued to
$K_{\mathbb{C}}.$

2) The map $\Pi_{\hbar}^{\ast}$ may be computed as
\[
\left(  \Pi_{\hbar}^{\ast}F\right)  \left(  x\right)  =\int_{\frak{k}}F\left(
xe^{iY}\right)  e^{-\left|  Y\right|  ^{2}/2\hbar}\zeta\left(  Y\right)  \,dY,
\]
where $\zeta$ is defined by (\ref{zeta.def}) and computed in Proposition
\ref{zeta.form} below.

3) There exists a constant $b_{\hbar}$ such that $b_{\hbar}\Pi_{\hbar}$ is a
unitary map of $L^{2}(K)$ onto $\mathcal{H}L^{2}(T^{\ast}(K),\gamma_{\hbar}).$
Thus $\Pi_{\hbar}^{\ast}=b_{\hbar}^{-2}\Pi_{\hbar}^{-1}.$

The constants $a_{\hbar}$ and $b_{\hbar}$ are given explicitly as $a_{\hbar
}=\left(  2\pi\hbar\right)  ^{n/2}e^{-\left|  \rho\right|  ^{2}\hbar/2}$ and
$b_{\hbar}=\left(  4\pi\hbar\right)  ^{-n/4},$ where $\rho$ is half the sum of
the positive roots for $K.$
\end{theorem}

\textit{Remarks}. 1) In the first expression for $\Pi_{\hbar}f,$ the analytic
continuation is in the space variable, from $K$ to $T^{\ast}(K)\cong
K_{\mathbb{C}},$ with $\hbar$ fixed. The map $\Pi_{\hbar}$ coincides (up to
the constant $a_{\hbar}$) with the generalized Segal--Bargmann transform for
$K,$ as described in \cite[Thm. 2]{H1}.

2) The formula for $\Pi_{\hbar}^{\ast}F$ may be taken literally on a dense
subspace of $\mathcal{H}L^{2}(T^{\ast}(K),\gamma_{\hbar}).$ For general $F,$
however, one should integrate over a ball of radius $R$ in $\frak{k}$ and then
take a limit in $L^{2}\left(  K\right)  ,$ as in \cite[Thm. 1]{H2}.

3) The formula for $\Pi_{\hbar}^{\ast}$ is an immediate consequence of the
formula (\ref{grppair2}) for the pairing. By computing $\zeta\left(  Y\right)
$ explicitly we may recognize $\Pi_{\hbar}^{\ast}$ as simply a constant times
the \textit{inverse Segal--Bargmann transform} for $K,$ as described in
\cite{H2}.

4) In \cite{H2} I deduce the unitarity of the generalized Segal--Bargmann
transform from the inversion formula. However, I do not know how to prove the
unitarity of the pairing map without recognizing that the measure in the
formula for $\Pi_{\hbar}^{\ast}$ is related to the heat kernel measure for
$K_{\mathbb{C}}/K.$

5) Since $F$ is holomorphic, there can be many different formulas for
$\Pi_{\hbar}^{\ast}$ (or $\Pi_{\hbar}^{-1}$). In particular, if one takes the
second expression for $\Pi_{\hbar}$ and computes the adjoint in the obvious
way, one will \textit{not} get the given expression for $\Pi_{\hbar}^{\ast}.$
Nevertheless, the two expressions for $\Pi_{\hbar}^{\ast}$ do agree on
holomorphic functions.

\textit{Proof}. We begin by writing the explicit formula for $\zeta.$

\begin{proposition}
\label{zeta.form}The function $\zeta$ is an Ad-$K$-invariant function on
$\frak{k}$ which is given on a maximal abelian subalgebra by
\[
\zeta\left(  Y\right)  =\prod_{\alpha\in R^{+}}\frac{\sinh\alpha(Y/2)}%
{\alpha(Y/2)},
\]
where $R^{+}$ is a system of positive roots.
\end{proposition}

The proof of this proposition is a straightforward but tedious calculation,
which I defer to an appendix.

Directly from the formula (\ref{grppair1}) for the pairing map we see that
\begin{equation}
\left\langle F,f\right\rangle _{pair}=\left\langle \Pi_{\hbar}^{\ast
}F,f\right\rangle _{L^{2}\left(  K\right)  },\label{pairmap1}%
\end{equation}
where $\Pi_{\hbar}^{\ast}$ is defined by
\[
\left(  \Pi_{\hbar}^{\ast}F\right)  \left(  x\right)  =\int_{\frak{k}}F\left(
xe^{iY}\right)  e^{-\left|  Y\right|  ^{2}/2\hbar}\zeta\left(  Y\right)  \,dY.
\]
At the moment it is not at all clear that $\Pi_{\hbar}$ is a bounded operator,
but there is a dense subspace of $\mathcal{H}L^{2}(T^{\ast}(K),\gamma_{\hbar
})$ on which $\Pi_{\hbar}$ makes sense and for which (\ref{pairmap1}) holds.
We will see below that $\Pi_{\hbar}$ extends to a bounded operator on all of
$\mathcal{H}L^{2}(T^{\ast}(K),\gamma_{\hbar}),$ for which (\ref{pairmap1})
continues to hold. Then by taking the adjoint of $\Pi_{\hbar}^{\ast}$ we see
that $\left\langle F,f\right\rangle _{pair}=\left\langle F,\Pi_{\hbar
}f\right\rangle _{\mathcal{H}L^{2}(T^{\ast}(K),\gamma_{\hbar})}$ as well.

Using the explicit formula for $\zeta$ and making the change of variable
$Y^{\prime}=\frac{1}{2}Y$ we have
\[
\left(  \Pi_{\hbar}^{\ast}F\right)  \left(  x\right)  =2^{n}\int_{\frak{k}%
}F\left(  xe^{2iY^{\prime}}\right)  \left[  e^{-2\left|  Y^{\prime}\right|
^{2}/\hbar}\prod_{\alpha\in R^{+}}\frac{\sinh\alpha\left(  Y^{\prime}\right)
}{\alpha\left(  Y^{\prime}\right)  }\,dY^{\prime}\right]  \,
\]
We recognize from \cite{H3} the expression in square brackets as a constant
times the heat kernel measure on $K_{\mathbb{C}}/K,$ written in exponential
coordinates and evaluated at time $t=\hbar/2.$ It follows from the inversion
formula of \cite{H2} that
\[
\Pi_{\hbar}^{\ast}=c_{\hbar}C_{\hbar}^{-1},
\]
for some constant $c_{\hbar}$ and where $C_{\hbar}$ is the generalized
Segal--Bargmann transform of \cite[Thm. 2]{H1}.

Now, $C_{\hbar}$ is unitary if we use on $K_{\mathbb{C}}\cong T^{\ast}(K)$ the
heat kernel measure $\nu_{\hbar}.$ But in Theorem 1 we established that this
measure coincides up to a constant with the measure $\gamma_{\hbar}.$ Thus
$\Pi_{\hbar}$ is a constant multiple of a unitary and coincides with
$C_{\hbar}$ up to a constant. This gives us what we want except for computing
the constants, which I leave as an exercise for the reader.$\,\square$

\section{Quantization, reduction, and Yang--Mills theory}

Let me summarize the results of this section before explaining them in detail.
It is possible to realize a compact Lie group $K$ as the quotient
$K=\mathcal{A}/\mathcal{L}\left(  K\right)  $ where $\mathcal{A}$ is a certain
infinite-dimensional Hilbert space and $\mathcal{L}\left(  K\right)  $ is the
based loop group over $K,$ which acts freely and isometrically on
$\mathcal{A}.$ (Here $\mathcal{A}$ is to be interpreted as a space of
connections over $S^{1}$ and $\mathcal{L}\left(  K\right)  $ as a gauge
group.) The cotangent bundle of $\mathcal{A}$ may be identified with the
associated complex Hilbert space $\mathcal{A}_{\mathbb{C}}$ and the symplectic
quotient $\mathcal{A}_{\mathbb{C}}//\mathcal{L}\left(  K\right)  $ is
identifiable with $T^{\ast}(K).$ The results of \cite{DH,Wr} (see also the
exposition in \cite{H8}) together with the results of this paper may be
interpreted as saying that in this case \textit{quantization commutes with
reduction}. This means two things. First, if we perform geometric quantization
on $\mathcal{A}_{\mathbb{C}}$ and then reduce by $\mathcal{L}\left(  K\right)
$ the resulting Hilbert space is naturally \textit{unitarily} equivalent to
the result of first reducing by $\mathcal{L}\left(  K\right)  $ and then
quantizing the reduced manifold $\mathcal{A}_{\mathbb{C}}//\mathcal{L}\left(
K\right)  =T^{\ast}(K).$ This result holds using either the vertical or the
K\"{a}hler polarization; in the K\"{a}hler case it is necessary to include the
half-form correction. Second, the pairing map between the vertically polarized
and K\"{a}hler-polarized Hilbert spaces over $\mathcal{A}_{\mathbb{C}}$
descends to the reduced Hilbert spaces and then coincides (up to a constant)
with the pairing map for $T^{\ast}(K).$ Additional discussion of these ideas
is found in \cite{H7,H8}. The first result contrasts with those of Guillemin
and Sternberg in \cite{GStern}. That paper considers the geometric
quantization of compact K\"{a}hler manifolds, without half-forms, and exhibits
(under suitable regularity assumptions) a one-to-one onto linear map between
the ``first quantize then reduce'' space and the ``first reduce and then
quantize'' space. However, they do not show that this map is unitary, and it
seems very unlikely that it is unitary in general. In the case considered in
this paper and \cite{DH}, quantization commutes \textit{unitarily} with reduction.

Consider then a Lie group $K$ of compact type, with a fixed Ad-$K$-invariant
inner product on its Lie algebra $\frak{k}.$ Then consider the real Hilbert
space
\[
\mathcal{A}:=L^{2}\left(  \left[  0,1\right]  ;\frak{k}\right)  .
\]
Let $\mathcal{L}\left(  K\right)  $ denote the \textit{based loop group} for
$K,$ namely the group of maps $l:\left[  0,1\right]  \rightarrow K$ such that
$l_{0}=l_{1}=e.$ (For technical reasons I\ also assume that $l$ has one
derivative in $L^{2},$ i.e. that $l$ has ``finite energy.'') There is a
natural action of $\mathcal{L}\left(  K\right)  $ on $\mathcal{A}$ given by
\begin{equation}
\left(  l\cdot A\right)  _{\tau}=l_{\tau}A_{\tau}l_{\tau}^{-1}-\frac{dl}%
{d\tau}l_{\tau}^{-1}.\label{gaugeact}%
\end{equation}
Here $l$ is in $\mathcal{L}\left(  K\right)  ,$ $A$ is in $\mathcal{A},$ and
$\tau$ is in $\left[  0,1\right]  .$ Then we have the following result: the
based loop group $\mathcal{L}\left(  K\right)  $ acts freely and isometrically
on $\mathcal{A},$ and the quotient $\mathcal{A}/\mathcal{L}\left(  K\right)  $
is a finite-dimensional manifold that is isometric to $K.$ Thus $K,$ which is
finite-dimensional but with non-trivial geometry, can be realized as a
quotient of $\mathcal{A}$, which is infinite-dimensional but flat.

Explicitly the quotient map is given in terms of the \textit{holonomy}. For
$A\in\mathcal{A}$ we define the holonomy $h\left(  A\right)  \in K$ by the
``path-ordered integral''
\begin{align}
h(A)  & =\mathcal{P}\left(  e^{\int_{0}^{1}A_{\tau}\,d\tau}\right) \nonumber\\
& =\lim_{N\rightarrow\infty}e^{\int_{0}^{1/N}A_{\tau}d\tau}e^{\int_{1/N}%
^{2/N}A_{\tau}d\tau}\cdots e^{\int_{(N-1)/N}^{1}A_{\tau}d\tau}%
.\label{holonomy}%
\end{align}
Then it may be shown that $A$ and $B$ are in the same orbit of $\mathcal{L}%
\left(  K\right)  $ if and only if $h(A)=h\left(  B\right)  .$ Furthermore,
every $x\in K$ is the holonomy of some $A\in\mathcal{A},$ and so the
$\mathcal{L}\left(  K\right)  $-orbits are in one-to-one correspondence with
points in $K.$ The motivation for these constructions comes from gauge theory.
The space $\mathcal{A}$ is to be thought of as the space of connections for a
trivial principal $K$-bundle over $S^{1},$ in which case $\mathcal{L}\left(
K\right)  $ is the based gauge group and (\ref{gaugeact}) is a gauge
transformation. For connections $A$ over $S^{1}$ the only quantity invariant
under (based) gauge transformations is the holonomy $h\left(  A\right)  $
around the circle. See \cite{DH} or \cite{H8} for further details.

Meanwhile, we may consider the cotangent bundle of $\mathcal{A},$ $T^{\ast
}\left(  \mathcal{A}\right)  ,$ which may be identified with
\[
\mathcal{A}_{\mathbb{C}}:=L^{2}\left(  \left[  0,1\right]  ;\frak{k}%
_{\mathbb{C}}\right)  .
\]
Then $\mathcal{A}_{\mathbb{C}}$ is an infinite-dimensional flat K\"{a}hler
manifold. The action of the based loop group $\mathcal{L}\left(  K\right)  $
on $\mathcal{A}$ extends in a natural way to an action on $\mathcal{A}%
_{\mathbb{C}}$ (given by the same formula). Starting with $\mathcal{A}%
_{\mathbb{C}}$ we may construct the symplectic (or Marsden-Weinstein) quotient
$\mathcal{A}_{\mathbb{C}}//\mathcal{L}\left(  K\right)  .$ This quotient is
naturally identifiable with $T^{\ast}\left(  \mathcal{A}/\mathcal{L}\left(
K\right)  \right)  =T^{\ast}(K).$ One may also realize the symplectic quotient
as $\mathcal{A}_{\mathbb{C}}/\mathcal{L}\left(  K_{\mathbb{C}}\right)  $ where
$\mathcal{L}\left(  K_{\mathbb{C}}\right)  $ is the based loop group over
$K_{\mathbb{C}}.$ The quotient $\mathcal{A}_{\mathbb{C}}/\mathcal{L}\left(
K_{\mathbb{C}}\right)  $ is naturally identifiable with $K_{\mathbb{C}}.$ So
we have ultimately
\[
T^{\ast}(K)\cong T^{\ast}\left(  \mathcal{A}/\mathcal{L}\left(  K\right)
\right)  \cong\mathcal{A}_{\mathbb{C}}/\mathcal{L}\left(  K_{\mathbb{C}%
}\right)  \cong K_{\mathbb{C}}.
\]
The resulting identification of $T^{\ast}(K)$ with $K_{\mathbb{C}}$ is nothing
but the one used throughout this paper. The quotient $\mathcal{A}_{\mathbb{C}%
}/\mathcal{L}\left(  K_{\mathbb{C}}\right)  $ may be expressed in terms of the
complex holonomy. For $Z\in\mathcal{A}_{\mathbb{C}}$ we define $h_{\mathbb{C}%
}\left(  Z\right)  \in K_{\mathbb{C}}$ similarly to (\ref{holonomy}). Then the
$\mathcal{L}\left(  K_{\mathbb{C}}\right)  $-orbits are labeled precisely by
the value of $h_{\mathbb{C}}.$

So the manifold $T^{\ast}(K)$ that we have been quantizing is a symplectic
quotient of the infinite-dimensional flat K\"{a}hler manifold $\mathcal{A}%
_{\mathbb{C}}.$ Looking at $T^{\ast}(K)$ in this way we may say that we have
first reduced $\mathcal{A}_{\mathbb{C}}$ by the loop group $\mathcal{L}\left(
K\right)  ,$ and then quantized. One may attempt to do things the other way
around: \textit{first} quantize $\mathcal{A}_{\mathbb{C}}$ and \textit{then}
reduce by $\mathcal{L}\left(  K\right)  .$ Motivated by the results of K. Wren
\cite{Wr} (see also \cite[Chap. IV.3.8]{La2}), Bruce Driver and I considered
precisely this procedure \cite{DH}. Although there are technicalities that
must be attended to in order to make sense of this, the upshot is that in this
case \textit{quantization commutes with reduction}, as explained in the first
paragraph of this section.

In the end we have three different procedures for constructing the generalized
Segal--Bargmann space for $K$ and the associated Segal--Bargmann transform.
The first is the heat kernel construction of \cite{H1}, the second is
geometric quantization of $T^{\ast}(K)$ with a K\"{a}hler polarization, and
the third is by reduction from $\mathcal{A}_{\mathbb{C}}.$ It is not obvious
\textit{a priori} that any two of these constructions should agree. That all
three agree is an apparent miracle that should be understood better. I expect
that if one replaces the compact group $K$ with some other class of Riemannian
manifolds, then these constructions will not agree.

Let me now explain how the quantization of $\mathcal{A}_{\mathbb{C}}$ and the
reduction by $\mathcal{L}\left(  K\right)  $ are done in \cite{DH}. (See also
the expository article \cite{H8}.) In the interests of conveying the main
ideas I will permit myself to gloss over various technical issues that are
dealt with carefully in \cite{DH}. Although \cite{DH} does not use the
language of geometric quantization, it can easily be reformulated in those
terms. Now, the constructions of geometric quantization are not directly
applicable in the infinite-dimensional setting. On the other hand,
$\mathcal{A}_{\mathbb{C}}$ is just a flat Hilbert space and there are by now
many techniques for dealing with its quantization. Driver and I want to first
perform quantization on $\mathbb{C}^{n}$ and then let $n$ tend to infinity. If
one performs geometric quantization on $\mathbb{C}^{n}$ with a K\"{a}hler
polarization and the half-form correction one gets $\mathcal{H}L^{2}%
(\mathbb{C}^{n},\nu_{\hbar})$ where
\[
d\nu_{\hbar}=e^{-(\operatorname{Im}z)^{2}/\hbar}\,dz.
\]
See Section 5 below.

In this form we cannot let the dimension go to infinity because the measure is
Gaussian only in the imaginary directions. So we introduce a regularization
parameter $s>\hbar/2$ and modify the measure to
\[
dM_{s,\hbar}=\left(  \pi r\right)  ^{-n/2}\left(  \pi\hbar\right)
^{-n/2}e^{-(\operatorname{Im}z)^{2}/\hbar}e^{-(\operatorname{Re}z)^{2}/r},
\]
where $r=2(s-\hbar/2).$ The constants are chosen so that $M_{s,\hbar}$ is a
probability measure. If one rescales $M_{s,\hbar}$ by a suitable function of
$s$ and then lets $s$ tend to infinity one recovers the measure $\nu_{\hbar}.$
Our Hilbert space is then just $\mathcal{H}L^{2}(\mathbb{C}^{n},M_{s,\hbar}).$
Now we can let the dimension tend to infinity, and we get
\[
\mathcal{H}L^{2}\left(  \overline{\mathcal{A}_{\mathbb{C}}},M_{s,\hbar
}\right)  ,
\]
where $M_{s,\hbar}$ is a Gaussian measure on a certain ``extension''
$\overline{\mathcal{A}_{\mathbb{C}}}$ of $\mathcal{A}_{\mathbb{C}}.$ (See
\cite[Sect. 4.1]{DH}.) This we think of as the (regularized) K\"{a}%
hler-polarized Hilbert space.

Our next task is to perform the reduction by $\mathcal{L}\left(  K\right)  ,$
which means looking for functions in $\mathcal{H}L^{2}(\overline
{\mathcal{A}_{\mathbb{C}}},M_{s,\hbar})$ that are ``invariant'' in the
appropriate sense under the action of $\mathcal{L}\left(  K\right)  .$ The
notion of invariance should itself come from geometric quantization, by
``quantizing'' the action of $\mathcal{L}\left(  K\right)  $ on $\mathcal{A}%
_{\mathbb{C}}.$ Note that $\mathcal{L}\left(  K\right)  $ acts on
$\mathcal{A}$ by a combination of rotations and translations; the action of
$\mathcal{L}\left(  K\right)  $ on $\mathcal{A}_{\mathbb{C}}$ is then induced
from its action on $\mathcal{A}.$ Let us revert temporarily to the
finite-dimensional situation as in Section 4. Then the way we have chosen our
1-form $\theta$ and our K\"{a}hler potential $\kappa$ means that the rotations
and translations of $\mathbb{R}^{n}$ act in the K\"{a}hler-polarized Hilbert
space $\mathcal{H}L^{2}(\mathbb{C}^{n},\nu_{\hbar})$ in the simplest possible
way, namely by rotating and translating the variables. (This is not the case
in the conventional form of the Segal--Bargmann space.) We will then formally
extend this notion to the infinite-dimensional case, which means that an
element $l$ of $\mathcal{L}\left(  K\right)  $ acts on a function
$F\in\mathcal{H}L^{2}(\overline{\mathcal{A}_{\mathbb{C}}},M_{s,\hbar})$ by
$F\left(  Z\right)  \rightarrow F\left(  l^{-1}\cdot Z\right)  .$

We want functions in $\mathcal{H}L^{2}(\overline{\mathcal{A}_{\mathbb{C}}%
},M_{s,\hbar})$ that are invariant under this action, i.e. such that $F\left(
l^{-1}\cdot Z\right)  =F\left(  Z\right)  $ for all $l\in\mathcal{L}\left(
K\right)  .$ Since our functions are holomorphic they must also (at least
formally) be invariant under $\mathcal{L}\left(  K_{\mathbb{C}}\right)  . $ So
we expect the invariant functions to be those of the form
\[
F\left(  Z\right)  =\Phi\left(  h_{\mathbb{C}}\left(  Z\right)  \right)
\]
where $\Phi$ is a holomorphic function on $K_{\mathbb{C}}.$ (Certainly every
such function is $\mathcal{L}\left(  K\right)  $-invariant. Although Driver
and I did not prove that every $\mathcal{L}\left(  K\right)  $-invariant
function is of this form, this is probably the case.) The norm of such a
function may be computed as
\[
\int_{\overline{\mathcal{A}_{\mathbb{C}}}}\left|  F\left(  Z\right)  \right|
^{2}dM_{s,\hbar}\left(  Z\right)  =\int_{K_{\mathbb{C}}}\left|  \Phi\left(
g\right)  \right|  ^{2}d\mu_{s,\hbar}\left(  g\right)
\]
where $\mu_{s,\hbar}$ is the push-forward of $M_{s,\hbar}$ to $K_{\mathbb{C}}$
under $h_{\mathbb{C}}.$ Concretely $\mu_{s,\hbar}$ is a certain heat kernel
measure on $K_{\mathbb{C}}.$ See \cite{DH} or \cite{H5} for details.

So our regularized reduced quantum Hilbert space is
\[
\mathcal{H}L^{2}(K_{\mathbb{C}},\mu_{s,\hbar})
\]
At this point we may remove the regularization by letting $s$ tend to
infinity. It can be shown that
\[
\lim_{s\rightarrow\infty}\mu_{s,\hbar}=\nu_{\hbar}%
\]
where $\nu_{\hbar}$ is the $K$-invariant heat kernel measure of \cite{H1}. So
without the regularization our reduced quantum Hilbert space becomes finally
\[
\mathcal{H}L^{2}(K_{\mathbb{C}},\nu_{\hbar}),
\]
which (up to a constant) is the same as $\mathcal{H}L^{2}(T^{\ast}%
(K),\gamma_{\hbar})$, using our identification of $T^{\ast}(K)$ with
$K_{\mathbb{C}}.$

Meanwhile the vertically polarized Hilbert space for $\mathbb{C}^{n}$ also
requires a regularization before we let $n$ tend to infinity. So we consider
$L^{2}(\mathbb{R}^{n},P_{s}),$ where $P_{s}$ is the Gaussian measure given by
\[
dP_{s}\left(  x\right)  =\left(  2\pi s\right)  ^{-n/2}e^{-\left|  x\right|
^{2}/2s}.
\]
Rescaling $P_{s}$ by a function of $s$ and then letting $s$ tend to infinity
gives back Lebesgue measure on $\mathbb{R}^{n}.$ We then consider the
Segal--Bargmann transform $S_{\hbar},$ which coincides with the pairing map of
geometric quantization (Section 5). This is given by
\[
S_{\hbar}f\left(  z\right)  =\left(  2\pi t\right)  ^{-n/2}\int_{\mathbb{R}%
^{n}}e^{-(z-x)^{2}/2t}f\left(  x\right)  \,dx.
\]
With the constants adjusted as above this map has the property that it is
unitary between our regularized spaces $L^{2}\left(  \mathbb{R}^{n}%
,P_{s}\right)  $ and $\mathcal{H}L^{2}(\mathbb{C}^{n},M_{s,\hbar}),$ for all
$s>\hbar/2.$ (See \cite[Sect. 3.1]{DH} or \cite{H5}.)

Letting the dimension tend to infinity we get a unitary map \cite[Sect.
4.1]{DH}
\begin{equation}
S_{\hbar}:L^{2}\left(  \overline{\mathcal{A}},P_{s}\right)  \rightarrow
\mathcal{H}L^{2}(\overline{\mathcal{A}_{\mathbb{C}}},M_{s,\hbar}).\label{ssh}%
\end{equation}
It seems reasonable to think of this as the infinite-dimensional regularized
version of the pairing map for $\mathcal{A}_{\mathbb{C}}.$ To reduce by
$\mathcal{L}\left(  K\right)  $ we consider functions in $L^{2}\left(
\overline{\mathcal{A}},P_{s}\right)  $ that are $\mathcal{L}\left(  K\right)
$-invariant. According to an important theorem of Gross \cite{G1} these are
(as expected) precisely those of the form
\begin{equation}
f\left(  A\right)  =\phi\left(  h\left(  A\right)  \right)  ,\label{phiform}%
\end{equation}
where $\phi$ is a function on $K.$ The norm of such a function is computed as
\[
\int_{\overline{\mathcal{A}}}\left|  f\left(  A\right)  \right|  ^{2}%
\,dP_{s}\left(  A\right)  =\int_{K}\left|  \phi\left(  x\right)  \right|
^{2}\,d\rho_{s}\left(  x\right)  .
\]
Thus with the vertical polarization our reduced Hilbert space becomes
$L^{2}\left(  K,\rho_{s}\right)  .$ Since
\[
\lim_{s\rightarrow\infty}d\rho_{s}\left(  x\right)  =dx
\]
we recover in the limit the vertically polarizes subspace for $K.$ (Compare
\cite{Go}.)

\begin{theorem}
\cite{DH} Consider the Segal--Bargmann transform $S_{s,\hbar}$ of (\ref{ssh}).
Then consider a function $f\in L^{2}\left(  \overline{\mathcal{A}}%
,P_{s}\right)  $ of the form $f\left(  A\right)  =\phi\left(  h\left(
A\right)  \right)  ,$ with $\phi$ a function on $K.$ Then
\[
\left(  S_{\hbar}f\right)  \left(  Z\right)  =\Phi\left(  h_{\mathbb{C}%
}\left(  Z\right)  \right)
\]
where $\Phi$ is the holomorphic function on $K_{\mathbb{C}}$ given by
\[
\Phi=\text{ analytic continuation of }e^{\hbar\Delta_{K}/2}\phi.
\]
Restricting $S_{\hbar}$ to the $\mathcal{L}\left(  K\right)  $-invariant
subspace and then letting $s\rightarrow\infty$ gives the unitary map
\[
C_{\hbar}:L^{2}\left(  K,dx\right)  \rightarrow\mathcal{H}L^{2}(K_{\mathbb{C}%
},\nu_{\hbar})
\]
given by $\phi\rightarrow$ analytic continuation of $e^{\hbar\Delta_{K}/2}\phi.$
\end{theorem}

If we restrict $S_{\hbar}$ to the $\mathcal{L}\left(  K\right)  $-invariant
subspace but keep $s$ finite, then we get a modified form of the
Segal--Bargmann transform for $K,$ a unitary map $L^{2}\left(  K,\rho
_{s}\right)  \rightarrow\mathcal{H}L^{2}(K_{\mathbb{C}},\mu_{s,\hbar}),$ still
given by $\phi\rightarrow$ analytic continuation of $e^{\hbar\Delta_{K}/2}%
\phi.$ This transform is examined from a purely finite-dimensional point of
view in \cite{H5}.

So if we accept the constructions of \cite{DH} as representing regularized
forms of the geometric quantization Hilbert spaces and pairing map, then we
have the following conclusions. First, the K\"{a}hler-polarized and vertically
polarized Hilbert spaces for $\mathcal{A}_{\mathbb{C}}$, after reducing by
$\mathcal{L}\left(  K\right)  $ and removing the regularization, are naturally
unitarily equivalent to the K\"{a}hler-polarized and vertically polarized
Hilbert spaces for $T^{\ast}(K)=\mathcal{A}_{\mathbb{C}}//\mathcal{L}\left(
K\right)  .$ (I am including the half-forms in the construction of the
K\"{a}hler-polarized Hilbert spaces.) Second, the pairing map for
$\mathcal{A}_{\mathbb{C}},$ after restricting to the $\mathcal{L}\left(
K\right)  $-invariant subspace and removing the regularization, coincides with
the pairing map for $T^{\ast}(K).$ Both of these statements are to be
understood ``up to a constant.''

\section{The geodesic flow and the heat equation}

This section describes how the complex polarization on $T^{\ast}(K)$ can be
obtained from the vertical polarization by means of the \textit{imaginary-time
geodesic flow}. This description is supposed to make the appearance of the
heat equation in the pairing map seem more natural. After all the heat
operator is nothing but the \textit{imaginary-time \textbf{quantized} geodesic
flow}. This point of view is due to T. Thiemann \cite{T1,T3}.

Suppose that $f$ is a function on $K$ and let $\pi:T^{\ast}(K)\rightarrow K $
be the projection map. Then $f\circ\pi$ is the extension of $f$ to $T^{\ast
}(K)$ that is constant along the fibers. A function of the form $f\circ\pi$ is
a ``vertically polarized function,'' that is, constant along the leaves of the
vertical polarization. Now recall the function $\kappa:T^{\ast}(K)\rightarrow
\mathbb{R}$ given by
\[
\kappa\left(  x,Y\right)  =\left|  Y\right|  ^{2}.
\]
Let $\Gamma_{t}$ be the Hamiltonian flow on $T^{\ast}(K)$ generated by the
function $\kappa/2.$ This is the geodesic flow for the bi-invariant metric on
$K$ determined by the inner product on the Lie algebra. The following result
gives a way of using the geodesic flow to produce a holomorphic function on
$T^{\ast}(K).$

\begin{theorem}
\label{igeod}Let $f:K\rightarrow\mathbb{C}$ be any function that admits an
entire analytic continuation to $T^{\ast}(K)\cong K_{\mathbb{C}},$ for
example, a finite linear combination of matrix entries. Let $\pi:T^{\ast
}(K)\rightarrow K$ be the projection map, and let $\Gamma_{t}$ be the geodesic
flow on $T^{\ast}(K).$

Then for each $m\in T^{\ast}(K)$ the map
\[
t\rightarrow f\left(  \pi\left(  \Gamma_{t}\left(  m\right)  \right)  \right)
\]
admits an entire analytic continuation (in $t$) from $\mathbb{R}$ to
$\mathbb{C}.$ Furthermore the function $f_{\mathbb{C}}:T^{\ast}(K)\rightarrow
\mathbb{C}$ given by
\[
f_{\mathbb{C}}\left(  m\right)  =f\left(  \pi\left(  \Gamma_{i}\left(
m\right)  \right)  \right)
\]
is holomorphic on $T^{\ast}(K)$ and agrees with $f$ on $K\subset T^{\ast}(K).$
\end{theorem}

Note that $f_{\mathbb{C}}$ is the analytic continuation of $f$ from $K$ to
$T^{\ast}(K)$, with respect to the complex structure on $T^{\ast}(K)$ obtained
by identifying it with $K_{\mathbb{C}}.$ So in words: to analytically continue
$f$ from $K$ to $T^{\ast}(K),$ first extend $f$ by making it constant along
the fibers and then compose with the time $i$ geodesic flow. So we can say
that the K\"{a}hler-polarized functions (i.e. holomorphic) are obtained from
the vertically polarized functions (i.e. constant along the fibers) by
composition with the time $i$ geodesic flow.

Now if $g$ is any function on $T^{\ast}(K)$ then $g\circ\Gamma_{t}$ may be
computed formally as
\[
g\circ\Gamma_{t}=\sum_{n=0}^{\infty}\frac{\left(  t/2\right)  ^{n}}%
{n!}\underset{n}{\underbrace{\left\{  \cdots\left\{  \left\{  g,\kappa
\right\}  ,\kappa\right\}  ,\cdots,\kappa\right\}  }}.
\]
Thus formally we have
\begin{equation}
f_{\mathbb{C}}=\sum_{n=0}^{\infty}\frac{\left(  i/2\right)  ^{n}}{n!}%
\underset{n}{\underbrace{\left\{  \cdots\left\{  \left\{  f\circ\pi
,\kappa\right\}  ,\kappa\right\}  ,\cdots,\kappa\right\}  }}.\label{ac.series}%
\end{equation}
(Compare \cite[Eq. (2.3)]{T1}.) In fact, this series converges provided only
that $f$ has an analytic continuation to $T^{\ast}(K).$ This series is the
``Taylor series in the fibers'' of $f_{\mathbb{C}}$; that is, on each fiber
the $n$th term of (\ref{ac.series}) is a homogeneous polynomial of degree $n. $

\begin{theorem}
\label{geoseries}Suppose $f$ is any function on $K$ that admits an entire
analytic continuation to $T^{\ast}(K),$ denoted $f_{\mathbb{C}}.$ Then the
series on the right in (\ref{ac.series}) converges absolutely at every point
and the sum is equal to $f_{\mathbb{C}}.$
\end{theorem}

As an illustrative example, consider the case $K=\mathbb{R}$ so that $T^{\ast
}(K)=\mathbb{R}^{2}.$ Then consider the function $f\left(  x\right)  =x^{k}$
on $\mathbb{R},$ so that $\left(  f\circ\pi\right)  \left(  x,y\right)
=x^{k}.$ Using the standard Poisson bracket on $\mathbb{R}^{2},$ $\left\{
g,h\right\}  =\frac{\partial g}{\partial x}\frac{\partial h}{\partial y}%
-\frac{\partial g}{\partial y}\frac{\partial h}{\partial x}$ it is easily
verified that
\[
\sum_{n=0}^{\infty}\frac{\left(  i/2\right)  ^{n}}{n!}\underset{n}%
{\underbrace{\left\{  \cdots\left\{  \left\{  x^{k},y^{2}\right\}
,y^{2}\right\}  ,\cdots,y^{2}\right\}  }}=\left(  x+iy\right)  ^{k}.
\]
(The series terminates after the $n=k$ term.) So $f_{\mathbb{C}}\left(
x+iy\right)  =\left(  x+iy\right)  ^{k}$ is indeed the analytic continuation
of $x^{k}.$

So ``classically'' the transition from the vertical polarization (functions
constant along the fibers) to the K\"{a}hler polarization (holomorphic
functions) is accomplished by means of the time $i$ geodesic flow. Let us then
consider the quantum counterpart of this, namely the transition from the
vertically polarized Hilbert space to the K\"{a}hler-polarized Hilbert space.
In the position Hilbert space the quantum counterpart of the function
$\kappa/2$ is the operator
\[
H:=-\hbar^{2}\Delta_{K}/2.
\]
(Possibly one should add an ``author-dependent'' multiple of the scalar
curvature to this operator \cite{O}, but since the scalar curvature of $K$ is
constant, this does not substantively affect the answer.) The quantum
counterpart of the geodesic flow is then the operator
\[
\hat{\Gamma}_{t}:=\exp\left(  itH/\hbar\right)
\]
and so the time $i$ quantized geodesic flow is represented by the operator
\[
\hat{\Gamma}_{i}=e^{\hbar\Delta_{K}/2}.
\]
Since this is precisely the heat operator for $K,$ the appearance of the heat
operator in the formula for the pairing map perhaps does not seem quite so
strange as at first glance.

This way of thinking about the complex structure and the associated
Segal--Bargmann transform is due to T. Thiemann \cite{T1}. The relationship
between the complex structure and the imaginary time geodesic flow is also
implicit in the work of Guillemin--Stenzel, motivated by the work of L. Boutet
de Monvel. (See the discussion between Thm. 5.2 and 5.3 in \cite{GStenz2}.)
Thiemann proposes a very general scheme for building complex structures and
Segal--Bargmann transforms (and their associated ``coherent states'') based on
these ideas. However, there are convergence issues that need to be resolved in
general, so it is not yet clear when one can carry this program out.

Although results similar to Theorems \ref{igeod} and \ref{geoseries} are
established in \cite[Lem. 3.1]{T3}, I give the proofs here for completeness.
Similar results hold for the ``adapted complex structure'' on the tangent
bundle of an real-analytic Riemannian manifold, which will be described elsewhere.

\textit{Proof}. According to a standard result \cite[Sect. IV.6]{He}, the
geodesics in $K$ are the curves of the form $\gamma\left(  t\right)  =xe^{tX},
$ with $x\in K$ and $X\in\frak{k}.$ This means that if we identify $T^{\ast
}(K)$ with $K\times\frak{k}$ by left-translation, then the geodesic flow takes
the form
\[
\Gamma_{t}\left(  x,Y\right)  =\left(  xe^{tY},Y\right)  .
\]
Thus if $f$ is a function on $K$ then
\[
f\left(  \pi\left(  \Gamma_{t}\left(  x,Y\right)  \right)  \right)  =f\left(
xe^{tY}\right)  .
\]

We are now supposed to fix $x$ and $Y$ and consider the map $t\rightarrow
f\left(  xe^{tY}\right)  .$ If $f$ has an analytic continuation to
$K_{\mathbb{C}},$ denoted $f_{\mathbb{C}},$ then the map $t\rightarrow
f\left(  xe^{tY}\right)  $ has an analytic continuation (in $t$) given by
\[
t\rightarrow f_{\mathbb{C}}\left(  xe^{tY}\right)  ,\quad t\in\mathbb{C}.
\]
(This because the exponential mapping from $\frak{k}_{\mathbb{C}}$ to
$K_{\mathbb{C}}$ is holomorphic.) Thus
\[
f\left(  \pi\left(  \Gamma_{i}\left(  x,Y\right)  \right)  \right)
=f_{\mathbb{C}}\left(  xe^{iY}\right)  .
\]
\qquad

Now we simply note that the map $\left(  x,Y\right)  \rightarrow
f_{\mathbb{C}}\left(  xe^{iY}\right)  $ is holomorphic on $T^{\ast}(K)$, with
respect to the complex structure obtained by the map $\Phi\left(  x,Y\right)
=xe^{iY}.$ This establishes Theorem \ref{igeod}.

To establish the series form of this result, Theorem \ref{geoseries}, we note
that (almost) by the definition of the geodesic flow we have
\begin{equation}
\left.  \left(  \frac{d}{dt}\right)  ^{n}\left(  f\circ\pi\right)  \circ
\Gamma_{t}\right|  _{t=0}=\frac{1}{2^{n}}\underset{n}{\underbrace{\left\{
\cdots\left\{  \left\{  f\circ\pi,\kappa\right\}  ,\kappa\right\}
,\cdots,\kappa\right\}  }}.\label{ftaylor}%
\end{equation}
On the other hand, if $f$ has an entire analytic continuation to $T^{\ast
}(K)\cong K_{\mathbb{C}},$ then as established above, the map $t\rightarrow
\left(  f\circ\pi\right)  \circ\Gamma_{t}$ has an entire analytic
continuation. This analytic continuation can be computed by an absolutely
convergent Taylor series at $t=0,$ where the Taylor coefficients at $t=0$ are
computable from (\ref{ftaylor}). Thus
\[
f_{\mathbb{C}}=\left(  f\circ\pi\right)  \circ\Gamma_{i}=\sum_{n=0}^{\infty
}\frac{(i/2)^{n}}{n!}\underset{n}{\underbrace{\left\{  \cdots\left\{  \left\{
f\circ\pi,\kappa\right\}  ,\kappa\right\}  ,\cdots,\kappa\right\}  }}.
\]
This establishes Theorem \ref{geoseries}.$\quad\square$

\section{The $\mathbb{R}^{n}$ case}

It is by now well known that geometric quantization can be used to construct
the Segal--Bargmann space for $\mathbb{C}^{n}$ and the associated
Segal--Bargmann transform. (See for example \cite[Sect. 9.5]{Wo}.) In this
section I repeat that construction, but in a manner that is non-standard in
two respects. First, I trivialize the quantum line bundle in such a way that
the measure in the Segal--Bargmann space is Gaussian only in the imaginary
directions. This is preferable for generalizing to the group case and it is a
simple matter in the $\mathbb{R}^{n}$ case to convert back to the standard
Segal--Bargmann space (see below). Second, I initially compute the pairing map
``backward,'' that is, from the Segal--Bargmann space to $L^{2}\left(
\mathbb{R}^{n}\right)  .$ I then describe this backward map in terms of the
backward heat equation, which leads to a description of the forward map in
terms of the forward heat equation. By contrast, Woodhouse uses the
reproducing kernel for the Segal--Bargmann space in order to compute the
pairing map in the forward direction. Although I include the half-form
correction on the complex side, this has no effect on the calculations in the
$\mathbb{R}^{n}$ case.

We consider the phase space $\mathbb{R}^{2n}=T^{\ast}\left(  \mathbb{R}%
^{n}\right)  $. We use the coordinates $q_{1},\cdots,q_{n},$ $p_{1}%
,\cdots,p_{n},$ where the $q$'s are the position variables and the $p$'s are
the momentum variables. We consider the \textit{canonical one form}
\[
\theta=\sum p_{k}\,dq_{k}%
\]
where here and in the following the sum ranges from $1$ to $n.$ Then
\[
\omega:=-d\theta=\sum dq_{k}\wedge dp_{k}%
\]
is the canonical 2-form. We consider a trivial complex line bundle
$L=\mathbb{R}^{2n}\times\mathbb{C}$ with a notion of covariant derivative
given by
\[
\nabla_{X}=X-\frac{1}{i\hbar}\theta\left(  X\right)  .
\]
Here $\nabla_{X}$ acts on smooth sections of $L,$ which we think of as smooth
functions on $\mathbb{R}^{2n}.$

The \textit{prequantum Hilbert space} is the space of sections of $L$ that are
square-integrable with respect to the canonical volume measure on
$\mathbb{R}^{2n}.$ The canonical volume measure is the one given by
integrating the \textit{Liouville volume form} defined as
\begin{align*}
\varepsilon & =\frac{1}{n!}\omega\wedge\cdots\wedge\omega\quad\text{(}n\text{
times)}\\
& =dq_{1}\wedge dp_{1}\wedge\cdots\wedge dq_{n}\wedge dp_{n}.
\end{align*}
Since our prequantum line bundle is trivial we may identify the prequantum
Hilbert space with $L^{2}\left(  \mathbb{R}^{2n},\varepsilon\right)  .$

We now consider the usual complex structure on $\mathbb{R}^{2n}=\mathbb{C}%
^{n}.$ We think of this complex structure as defining a \textit{K\"{a}hler
polarization} on $\mathbb{R}^{2n}.$ This means that we define a smooth section
$s$ of $L$ to be \textit{polarized} if
\begin{equation}
\nabla_{\partial/\partial\bar{z}_{k}}s=0\label{polarized1}%
\end{equation}
for all $k.$

\begin{proposition}
If we think of sections $s$ of $L$ as functions then a smooth section $s$
satisfies (\ref{polarized1}) if and only if $s$ is of the form
\begin{equation}
s\left(  q,p\right)  =F\left(  q_{1}+ip_{1},\cdots,q_{n}+ip_{n}\right)
e^{-p^{2}/2\hbar}\label{polarized2}%
\end{equation}
where $F$ is a holomorphic function on $\mathbb{C}^{n}.$ Here $p^{2}=p_{1}%
^{2}+\cdots+p_{n}^{2}.$
\end{proposition}

\textit{Proof}. To prove this we first compute $\nabla_{\partial/\partial
\bar{z}_{k}}$ as
\begin{align*}
\nabla_{\partial/\partial\bar{z}_{k}}  & =\frac{\partial}{\partial\bar{z}_{k}%
}-\frac{1}{i\hbar}\theta\left(  \frac{\partial}{\partial\bar{z}_{k}}\right) \\
& =\frac{1}{2}\left(  \frac{\partial}{\partial q_{k}}+i\frac{\partial
}{\partial p_{k}}\right)  -\frac{1}{2i\hbar}p_{k}.
\end{align*}
Then we note that
\begin{align*}
\nabla_{\partial/\partial\bar{z}_{k}}e^{-p^{2}/2\hbar}  & =\left[  \frac{1}%
{2}\left(  \frac{\partial}{\partial q_{k}}+i\frac{\partial}{\partial p_{k}%
}\right)  \left(  -\frac{p^{2}}{2\hbar}\right)  -\frac{1}{2i\hbar}%
p_{k}\right]  e^{-p^{2}/2\hbar}\\
& =\left[  -i\frac{p_{k}}{2\hbar}-\frac{1}{2i\hbar}p_{k}\right]
e^{-p^{2}/2\hbar}\\
& =0.
\end{align*}
Then if $s$ is any section, we can write $s$ in the form $s=F\,e^{-p^{2}%
/2\hbar},$ for some complex-valued function $F.$ Such a section $s$ is
polarized if and only if
\begin{align*}
0  & =\nabla_{\partial/\partial\bar{z}_{k}}\left(  F\,e^{-p^{2}/2\hbar}\right)
\\
& =\frac{\partial F}{\partial\bar{z}_{k}}e^{-p^{2}/2\hbar}+F\,\nabla
_{\partial/\partial\bar{z}_{k}}e^{-p^{2}/2\hbar}\\
& =\frac{\partial F}{\partial\bar{z}_{k}}e^{-p^{2}/2\hbar},
\end{align*}
for all $k,$ that is, if and only if $F$ is holomorphic.$\,\square$

We then define the \textit{K\"{a}hler-polarized Hilbert space} to be the space
of square-integrable K\"{a}hler-polarized sections of $L.$ Note that the
$L^{2}$ norm of the section $s$ in (\ref{polarized2}) is computable as
\[
\left\|  s\right\|  ^{2}=\int_{\mathbb{C}^{n}}\left|  F\left(  z\right)
\right|  ^{2}e^{-p^{2}/\hbar}\,d^{n}q\,d^{n}p,
\]
where $z=q+ip$ with $q,p\in\mathbb{R}^{n}.$ If we identify the polarized
section $s$ with the holomorphic function $F$ then we identify the
K\"{a}hler-polarized Hilbert space as the space
\begin{equation}
\mathcal{H}L^{2}(\mathbb{C}^{n},e^{-p^{2}/\hbar}d^{n}q\,d^{n}p).\label{sb1}%
\end{equation}
Here $\mathcal{H}L^{2}$ denotes the space of square-integrable holomorphic
functions with respect to the indicated measure. This space is a form of the
\textit{Segal--Bargmann space}.

The conventional description \cite[Sect. 9.2]{Wo} of the Segal--Bargmann space
is slightly different from what we have here, for two reasons. First, it is
conventional to insert a factor of $\sqrt{2}$ into the identification of
$\mathbb{R}^{2n}$ with $\mathbb{C}^{n}.$ Second, it is common to use a
different trivialization of $L$, resulting in a different Gaussian measure on
$\mathbb{C}^{n}.$ The map $F\rightarrow e^{z^{2}/4\hbar}F$ maps ``my''
Segal--Bargmann space unitarily to $\mathcal{H}L^{2}(\mathbb{C}^{n}%
,e^{-\left|  z\right|  ^{2}/2\hbar}d^{n}q\,d^{n}p),$ which is the standard
Segal--Bargmann space (apart from the above-mentioned factor of $\sqrt{2}$).
The normalization used here for the $\mathbb{R}^{n}$ case is the one that
generalizes to the group case.

We also define the \textit{canonical bundle} (relative to the given complex
structure) to be the bundle whose sections are $n$-forms of type $\left(
n,0\right)  .$ We then define the \textit{half-form bundle} $\delta_{1}$ to be
the square root of the canonical bundle. The polarized sections of $\delta
_{1}$ are objects of the form
\[
F\left(  z\right)  \sqrt{dz_{1}\wedge\cdots\wedge dz_{n}},
\]
where $F$ is holomorphic. Here the square root is a mnemonic for a polarized
section of $\delta_{1}$ whose square is $dz_{1}\wedge\cdots\wedge dz_{n}.$ The
absolute value of such a section is computed by setting
\begin{align}
\left|  \sqrt{dz_{1}\wedge\cdots\wedge dz_{n}}\right|  ^{2}  & =\left[
\frac{d\bar{z}_{1}\wedge\cdots\wedge d\bar{z}_{n}\wedge dz_{1}\wedge
\cdots\wedge dz_{n}}{b\,\varepsilon}\right]  ^{1/2}\nonumber\\
& =1,\label{dzs}%
\end{align}
where the constant $b$ is given by $b=(-1)^{n(n-1)/2}(2i)^{n}.$

The \textit{half-form-corrected Hilbert space} is then the space of
square-integrable polarized sections of $L\otimes\delta_{1}.$ Polarized
sections of $L\otimes\delta_{1}$ may be expressed uniquely as
\begin{equation}
s=F\left(  z\right)  e^{-p^{2}/2\hbar}\otimes\sqrt{dz_{1}\wedge\cdots\wedge
dz_{n}}.\label{sbspace1}%
\end{equation}
In light of (\ref{dzs}) our Hilbert space may again be identified with the
Segal--Bargmann space $\mathcal{H}L^{2}(\mathbb{C}^{n},e^{-p^{2}/\hbar}%
\,d^{n}q\,d^{n}p).$ Although in this flat case the half-form correction does
not affect in the description of the Hilbert space, it still has an important
effect on certain subsequent calculations, such as the WKB approximation. (See
\cite[Chap. 10]{Wo}.)

Next we consider the \textit{vertically polarized sections}. A vertically
polarized section $s$ of $L$ is one for which $\nabla_{\partial/\partial
p_{k}}s=0$ for all $k.$ Identifying sections with functions and using
$\theta=\Sigma p_{k}dq_{k}$ we see that $\nabla_{\partial/\partial p_{k}%
}=\partial/\partial p_{k}.$ Thus the vertically polarized sections are simply
functions $f\left(  q,p\right)  $ that are independent of $p.$ Unfortunately,
such a section cannot be square-integrable (over $\mathbb{R}^{2n} $) unless it
is zero almost everywhere.

So we now consider the \textit{canonical bundle} (relative to the vertical
polarization). This is the \textit{real} line bundle whose sections are
$n$-forms $\alpha$ satisfying $(\partial/\partial p_{k})\lrcorner\alpha=0$ for
all $k.$ Concretely such forms are precisely those expressible as
\[
\alpha=f\left(  q,p\right)  \,dq_{1}\wedge\cdots\wedge dq_{n}%
\]
where $f$ is real-valued. Such a $n$-form is called \textit{polarized} if
$(\partial/\partial p_{k})\lrcorner d\alpha=0$ for all $k.$ Such forms are
precisely those expressible as
\[
\alpha=f\left(  q\right)  \,dq_{1}\wedge\cdots\wedge dq_{n}.
\]
We now choose an orientation on $\mathbb{R}^{n}$ and we construct a square
root $\delta_{2}$ of the canonical bundle in such a way that the square of a
section of $\delta_{2}$ is a non-negative multiple of $dq_{1}\wedge
\cdots\wedge dq_{n},$ where $q_{1},\cdots,q_{n}$ is an oriented coordinate
system for $\mathbb{R}^{n}.$ There is a natural notion of polarized sections
of $\delta_{2},$ namely those whose squares are polarized sections of the
canonical bundle. The polarized sections of $\delta_{2}$ are precisely those
of the form
\begin{equation}
\beta=f\left(  q\right)  \sqrt{dq_{1}\wedge\cdots\wedge dq_{n}}%
.\label{vert.form}%
\end{equation}

We then consider the space of polarized sections of $L\otimes\delta_{2}.$
Every such section may be written uniquely in the form
\begin{equation}
s=f\left(  q\right)  \otimes\sqrt{dq_{1}\wedge\cdots\wedge dq_{n}%
},\label{vertspace1}%
\end{equation}
where now $f$ is complex-valued. We define the inner product of two such
sections $s_{1}$ and $s_{2}$ by
\begin{equation}
\left(  s_{1},s_{2}\right)  =\int_{\mathbb{R}^{n}}\overline{f_{1}}\left(
q\right)  f_{2}\left(  q\right)  \,dq_{1}\wedge\cdots\wedge dq_{n}%
.\label{inner1}%
\end{equation}
Note that the integration is over $\mathbb{R}^{n}$ not $\mathbb{R}^{2n}.$ The
\textit{vertically polarized Hilbert space} is the space of polarized sections
$s$ of $L\otimes\delta_{2}$ for which $\left(  s,s\right)  <\infty. $ (This
construction is explained in a more manifestly coordinate-independent way in
the general group case, in Section 2.4.)

Finally, we introduce the \textit{pairing map} between the vertically
polarized and K\"{a}hler-polarized Hilbert spaces. First we define a pointwise
pairing between sections of $\delta_{1}$ and sections of $\delta_{2}$ by
setting
\begin{align*}
\left(  \sqrt{dz_{1}\wedge\cdots\wedge dz_{n}},\sqrt{dq_{1}\wedge\cdots\wedge
dq_{n}}\right)   & =\left[  \frac{d\bar{z}_{1}\wedge\cdots\wedge d\bar{z}%
_{n}\wedge dq_{1}\wedge\cdots\wedge dq_{n}}{c\,\varepsilon}\right]  ^{1/2}\\
& =1,
\end{align*}
where the constant $c$ is given by $c=(-i)^{n}(-1)^{n(n+1)/2}.$ Then we may
pair a section of $L\otimes\delta_{1}$ with a section of $L\otimes\delta_{2}$
by applying the above pairing of $\delta_{1}$ and $\delta_{2}$ and the
Hermitian structure on $L$, and then integrating with respect to
$\varepsilon.$ So if $s_{1}$ is a polarized section of $L\otimes\delta_{1}$ as
in (\ref{sbspace1}) and $s_{2}$ is a polarized section of $L\otimes\delta_{2}$
then we have explicitly
\begin{equation}
\left\langle F,f\right\rangle _{pair}=\int_{\mathbb{R}^{n}}\int_{\mathbb{R}%
^{n}}\overline{F\left(  q+ip\right)  }f\left(  q\right)  e^{-p^{2}/2\hbar
}\,d^{n}q\,d^{n}p.\label{pairing1}%
\end{equation}
Here I have expressed things in terms of $F\in\mathcal{H}L^{2}(\mathbb{C}%
^{n},e^{-p^{2}/\hbar}\,d^{n}q\,d^{n}p)$ and $f\in L^{2}(\mathbb{R}^{n}).$

\begin{theorem}
Let us identify the vertically polarized Hilbert space with $L^{2}\left(
\mathbb{R}^{n}\right)  $ as in (\ref{inner1}) and the K\"{a}hler-polarized
Hilbert space with $\mathcal{H}L^{2}(\mathbb{C}^{n},e^{-p^{2}/\hbar}%
d^{n}q\,d^{n}p)$ as in (\ref{sbspace1}). Then there exists a unique bounded
linear operator $\Pi_{\hbar}:L^{2}(\mathbb{R}^{n})\rightarrow\mathcal{H}%
L^{2}(\mathbb{C}^{n},e^{-p^{2}/\hbar}d^{n}q\,d^{n}p)$ such that
\[
\left\langle f,F\right\rangle =\left\langle \Pi_{\hbar}f,F\right\rangle
_{\mathcal{H}L^{2}(\mathbb{C}^{n},e^{-p^{2}/\hbar}\,d^{n}q\,d^{n}%
p)}=\left\langle f,\Pi_{\hbar}^{\ast}F\right\rangle _{L^{2}(\mathbb{R}^{n})}.
\]
We call $\Pi_{\hbar}$ the \textbf{pairing map}. We then have the following results.

1) The map $\Pi_{\hbar}:L^{2}(\mathbb{R}^{n})\rightarrow\mathcal{H}%
L^{2}(\mathbb{C}^{n},e^{-p^{2}/\hbar}d^{n}q\,d^{n}p)$ is given by
\[
\Pi_{\hbar}f\left(  z\right)  =a_{\hbar}\int_{\mathbb{R}^{n}}e^{-\left(
z-q\right)  ^{2}/2\hbar}f\left(  q\right)  \,d^{n}q
\]
where $a_{\hbar}=\left(  \pi\hbar\right)  ^{-n/2}\left(  2\pi\hbar\right)
^{-n}.$

2) The map $\Pi_{\hbar}^{\ast}$ may be computed as
\[
\left(  \Pi_{\hbar}^{\ast}F\right)  \left(  q\right)  =\int_{\mathbb{R}^{n}%
}F\left(  q+ip\right)  e^{-p^{2}/2\hbar}\,d^{n}p.
\]

3) The map $b_{\hbar}\Pi_{\hbar}$ is unitary, where $b_{\hbar}=\left(
\pi\hbar\right)  ^{n/4}\left(  2\pi\hbar\right)  ^{n/2}.$
\end{theorem}

Note that the formula for $\Pi_{\hbar}^{\ast}$ (mapping from the
Segal--Bargmann space to $L^{2}(\mathbb{R}^{n})$) comes almost directly from
the formula (\ref{pairing1}) for the pairing. The unitarity (up to a constant)
of the pairing map in this $\mathbb{R}^{n}$ case is ``explained'' by the
Stone--von Neumann theorem. The map $\Pi_{\hbar}$, as given in 1), is the
``invariant'' form of the Segal--Bargmann transform, as described, for
example, in \cite[Sect. 6.3]{H6}. In the expression for $\Pi_{\hbar}^{\ast}$
the integral is not absolutely convergent in general, so more precisely one
should integrate over the set $\left|  p\right|  \leq R$ and then take a limit
(in $L^{2}(\mathbb{R}^{n})$) as $R\rightarrow\infty.$ (Compare \cite[Thm.
1]{H2}.)

There are doubtless many ways of proving these results. I will explain here
simply how the heat equation creeps into the argument, since the heat equation
is essential to the proof in the group case. Fix a holomorphic function $F$ on
$\mathbb{C}^{n}$ that is square-integrable over $\mathbb{R}^{n}$ and that has
moderate growth in the imaginary directions. Then define a function $f_{\hbar
}$ on $\mathbb{R}^{n}$ by
\begin{equation}
f_{\hbar}\left(  q\right)  =\int_{\mathbb{R}^{n}}F\left(  q+ip\right)  \left[
\frac{e^{-p^{2}/2\hbar}}{\left(  2\pi\hbar\right)  ^{n/2}}\right]
\,d^{n}p.\label{back1}%
\end{equation}
Note that the Gaussian factor in the square brackets is just the standard heat
kernel in the $p$-variable and in particular satisfies the forward heat
equation $\partial u/\partial\hbar=(1/2)\Delta u.$ Let us then differentiate
under the integral sign, integrate by parts, and use the Cauchy--Riemann
equations in the form $\partial F/\partial p_{k}=i\partial F/\partial q_{k}.$
This shows that
\begin{equation}
\frac{\partial f_{\hbar}}{\partial\hbar}=-\frac{1}{2}\Delta f_{\hbar
},\label{back2}%
\end{equation}
which is the \textit{backward} heat equation. Furthermore, letting $\hbar$
tend to zero we see that
\begin{equation}
\lim_{\hbar\downarrow0}f_{\hbar}\left(  q\right)  =F\left(  q\right)
.\label{back3}%
\end{equation}

Thus (up to a factor of $\left(  2\pi\hbar\right)  ^{n/2}$) $\Pi_{\hbar}%
^{\ast}F$ is obtained by applying the \textit{inverse} heat operator to the
restriction of $F$ to $\mathbb{R}^{n}.$ Turning this the other way around we
have
\begin{equation}
\left(  \Pi_{\hbar}^{\ast}\right)  ^{-1}f=\text{ }\left(  2\pi\hbar\right)
^{n/2}\left(  \text{analytic continuation of }e^{\hbar\Delta/2}f\right)
\label{pistarinv}%
\end{equation}
where $e^{\hbar\Delta/2}f$ means the solution to the heat operator at time
$\hbar,$ with initial condition $f.$ Of course, $e^{\hbar\Delta/2}f$ can be
computed by integrating $f$ against a Gaussian, so we have
\[
\left(  \Pi_{\hbar}^{\ast}\right)  ^{-1}f\left(  z\right)  =\int
_{\mathbb{R}^{n}}e^{-(z-q)^{2}/2\hbar}f\left(  q\right)  \,d^{n}q
\]
where the factors of $2\pi\hbar$ in (\ref{pistarinv}) have canceled those in
the computation of the heat operator on $\mathbb{R}^{n}.$

We now recognize $\left(  \Pi_{\hbar}^{\ast}\right)  ^{-1}$ as coinciding up
to a constant with the ``invariant'' form $C_{\hbar}$ of the Segal--Bargmann
transform, as described in \cite[Sect. 6.3]{H6}. The unitarity of $C_{\hbar}$
then implies that $\Pi_{\hbar}$ is unitary up to a constant. The argument in
the compact group case goes in much the same way, using the inversion formula
\cite{H2} for the generalized Segal--Bargmann transform of \cite{H1}.

\section{Appendix: Calculations with $\zeta$ and $\kappa$}

We will as always identify $T^{\ast}(K)$ with $K\times\frak{k}$ by means of
left-translation and the inner product on $\frak{k}.$ We choose an orthonormal
basis for $\frak{k}$ and we let $y_{1},\cdots,y_{n}$ be the coordinates with
respect to this basis. Then all forms on $K\times\frak{k}$ can be expressed in
terms of the left-invariant 1-forms $\eta_{1},\cdots,\eta_{n}$ on $K$ and the
translation-invariant 1-forms $dy_{1},\cdots,dy_{n}$ on $\frak{k}.$ Since the
canonical projection $pr:T^{\ast}(K)\rightarrow K$ in this description is just
projection onto the $K$ factor, $pr^{\ast}\left(  \eta_{k}\right)  $ is just
identified with $\eta_{k}.$ We identify the tangent space at each point in
$K\times\frak{k}$ with $\frak{k}+\frak{k}.$

Meanwhile we identify the tangent space of $K_{\mathbb{C}}$ at each point with
$\frak{k}_{\mathbb{C}}\cong\frak{k}+\frak{k}.$ We then consider the map $\Phi$
that identifies $T^{\ast}(K)\cong K\times\frak{k}$ with $K_{\mathbb{C}},$%
\[
\Phi\left(  x,Y\right)  =xe^{iY}.
\]
Since we are identifying the tangent space at every point of both
$K\times\frak{k}$ and $K_{\mathbb{C}}$ with $\frak{k}+\frak{k},$ the
differential of $\Phi$ at any point will be described as a linear map of
$\frak{k}+\frak{k}$ to itself. Explicitly we have \cite[Eq. (14)]{H3} at each
point $\left(  x,Y\right)  $
\begin{equation}
\Phi_{\ast}=\left(
\begin{array}
[c]{cc}%
\cos adY & \frac{1-\cos adY}{adY}\\
-\sin adY & \frac{\sin adY}{adY}%
\end{array}
\right)  .\label{phistar}%
\end{equation}

Our first task is to compute the function $\zeta(Y)$ defined in
(\ref{zeta.def}). So let us use $\Phi$ to pull back the left-invariant
anti-holomorphic forms $\bar{Z}_{k}$ to $T^{\ast}(K).$ To do this we compute
the adjoint $\Phi^{\ast}$ of the matrix (\ref{phistar}), keeping in mind that
$adY$ is skew, since our inner product is Ad-$K$-invariant. We then get that
\begin{align*}
\Phi^{\ast}\left(  \bar{Z}_{k}\right)   & =\text{ terms involving }\eta_{l}\\
& -i\left[  \frac{\sin adY}{adY}+i\frac{\cos adY-1}{adY}\right]  _{lk}dy_{l}.
\end{align*}
Thus
\begin{align*}
\bar{Z}_{1}\wedge\cdots\wedge\bar{Z}_{n}\wedge\eta_{1}\wedge\cdots\wedge
\eta_{n}  & =(-i)^{n}\zeta\left(  Y\right)  ^{2}\eta_{1}\wedge\cdots\wedge
\eta_{n}\wedge dy_{1}\wedge\cdots\wedge dy_{n}\\
& =\pm(-i)^{n}\zeta\left(  Y\right)  ^{2}\varepsilon
\end{align*}
where
\[
\zeta\left(  Y\right)  ^{2}=\det\left[  \frac{\sin adY}{adY}+i\frac{\cos
adY-1}{adY}\right]  .
\]
Here $\varepsilon=\eta_{1}\wedge dy_{1}\wedge\cdots\wedge\eta_{n}\wedge
dy_{n}$ is the Liouville volume form, and the factor of $\pm(-i)^{n}$ is
accounted for by the constant $c$ in the definition of $\zeta.$

Computing in terms of the roots we have
\begin{align*}
\zeta\left(  Y\right)  ^{2}  & =\prod_{\alpha\in R}\frac{\sinh\alpha
(Y)+\cosh\alpha(Y)-1}{\alpha(Y)}\\
& =\prod_{\alpha\in R}\frac{e^{\alpha(Y)}-1}{\alpha(Y)}\\
& =\prod_{\alpha\in R^{+}}\frac{\left(  e^{\alpha(Y)}-1\right)  \left(
1-e^{-\alpha(Y)}\right)  }{\alpha(Y)^{2}}%
\end{align*}
Since $\left(  e^{x}-1\right)  \left(  1-e^{-x}\right)  =4\sinh^{2}(x/2)$ we
get
\[
\zeta\left(  Y\right)  ^{2}=\prod_{\alpha\in R^{+}}\frac{\sinh^{2}\alpha
(Y/2)}{\alpha(Y/2)^{2}}.
\]
Taking a square root gives the desired expression for $\zeta(Y).$

Now we turn to the K\"{a}hler potential $\kappa.$ As usual we identify
$T^{\ast}(K)$ with $K\times\frak{k}$ by means of left-translation and the
inner product on $\frak{k}.$ The canonical projection $\pi:T^{\ast
}(K)\rightarrow K$ in this description is simply the map $\left(  x,Y\right)
\rightarrow x.$ The canonical 1-form $\theta$ is defined by setting
\[
\theta\left(  X\right)  =\left\langle Y,\pi_{\ast}\left(  X\right)
\right\rangle ,
\]
where $X$ is a tangent vector to $T^{\ast}(K)$ at the point $\left(
x,Y\right)  .$ Choose an orthonormal basis $e_{1},\cdots,e_{n}$ for $\frak{k}
$ and let $y_{1},\cdots,y_{n}$ be the coordinates on $\frak{k}$ with respect
to this basis. Let $\alpha_{1},\cdots,\alpha_{n}$ be left-invariant 1-forms on
$K$ whose values at the identity are the vectors $e_{1},\cdots,e_{n}$ in
$\frak{k}\cong\frak{k}^{\ast}.$ Then it is easily verified that at each point
$\left(  x,Y\right)  \in T^{\ast}(K)$ we have
\[
\theta=\sum_{k=1}^{n}y_{k}\alpha_{k}.
\]

Now let $\kappa$ be the function on $T^{\ast}(K)$ given by
\[
\kappa\left(  x,Y\right)  =\left|  Y\right|  ^{2}=\sum_{k=1}^{n}y_{k}^{2}.
\]
We want to verify that
\[
\operatorname{Im}\left[  \bar{\partial}\kappa\right]  =\theta.
\]
We start by observing that
\[
d\kappa=\sum_{k=1}^{n}2y_{k}\,dy_{k}.
\]

To compute $\bar{\partial}\kappa$ we need transport $d\kappa$ to
$K_{\mathbb{C}},$ where the complex structure is defined. On $K_{\mathbb{C}}$
we express things in terms of left-invariant 1-forms $\eta_{1},\cdots,\eta
_{n}$ and $J\eta_{1},\cdots,J\eta_{n}.$ We then want to pull back $d\kappa$ to
$K_{\mathbb{C}}$ by means of $\Phi^{-1}.$ So we need to compute the inverse
transpose of the matrix (\ref{phistar}) describing $\Phi_{\ast}$. This may be
computed as
\[
\left(  \Phi_{\ast}^{-1}\right)  ^{tr}=\frac{adY}{\sin adY}\left(
\begin{array}
[c]{cc}%
\frac{\sin adY}{adY} & -\sin adY\\
\frac{1-\cos adY}{adY} & \cos adY
\end{array}
\right)  .
\]
In terms of our basis for 1-forms on $T^{\ast}(K),$ $d\kappa$ is represented
by the vector
\[
\left[
\begin{array}
[c]{c}%
0\\
Y
\end{array}
\right]
\]
so we have to apply the matrix above to this vector. But of course $adY\left(
Y\right)  =0$ and so we get simply
\begin{align*}
\left(  \Phi^{-1}\right)  ^{\ast}\left(  d\kappa\right)   & =2\sum_{k=1}%
^{n}y_{k}J\eta_{k}\\
& =2\sum_{k=1}^{n}y_{k}\frac{1}{2i}\left(  \left(  \eta_{k}+iJ\eta_{k}\right)
-\left(  \eta_{k}-iJ\eta_{k}\right)  \right)  .
\end{align*}
Thus taking only the term involving the anti-holomorphic 1-forms $\eta
_{k}-iJ\eta_{k}$ we have
\[
\bar{\partial}\kappa=\sum_{k=1}^{n}iy_{k}\left(  \eta_{k}-iJ\eta_{k}\right)  .
\]
which is represented by the vector
\[
\left[
\begin{array}
[c]{c}%
iY\\
Y
\end{array}
\right]
\]

We now transfer this back to $T^{\ast}(K)$ by means of $\Phi^{\ast}.$ So
applying the transpose of the matrix (\ref{phistar}) we get
\[
\bar{\partial}\kappa=\sum_{k=1}^{n}\left(  iy_{k}\alpha_{k}+y_{k}%
dy_{k}\right)
\]
and so
\[
\operatorname{Im}\left[  \bar{\partial}\kappa\right]  =\sum_{k=1}^{n}%
y_{k}\alpha_{k}=\theta.
\]

\section{Appendix: Lie groups of compact type}

In this appendix I give a proof of Proposition \ref{structure.prop}, the
structure result for connected Lie groups of compact type. We consider a
connected Lie group $K$ of compact type, with a fixed Ad-invariant inner
product on its Lie algebra $\frak{k}.$ Since the inner product is
Ad-invariant, the orthogonal complement of any ideal in $\frak{k}$ will be an
ideal. Thus $\frak{k}$ decomposes as a direct sum of subalgebras that are
either simple or one-dimensional. Collecting together the simple factors in
one group and the one-dimensional factors in another, we obtain a
decomposition of $\frak{k}$ as $\frak{k}=\frak{k}_{1}+\frak{z},$ where
$\frak{k}_{1}$ is semisimple and $\frak{z}$ is commutative. Since
$\frak{k}_{1}$ is semisimple and admits an Ad-invariant inner product, the
connected subgroup $K_{1}$ of $K$ with Lie algebra $\frak{k}_{1}$ will be
compact. (By Cor. II.6.5 of \cite{He}, the adjoint group of $K_{1}$ is a
closed subgroup of $Gl\left(  \frak{k}_{1}\right)  \cap O\left(  \frak{k}%
_{1}\right)  $ and is therefore compact. Then Thm. II.6.9 of \cite{He} implies
that $K_{1}$ itself is compact.)

Now let $\Gamma$ be the subset of $\frak{z}$ given by
\[
\Gamma=\left\{  \left.  Z\in\frak{z}\right|  e^{Z}=id\right\}  ,
\]
where $id$ is the identity in $K.$ Since $\frak{z}$ is commutative, $\Gamma$
is a discrete additive subgroup of $\frak{z},$ hence there exist vectors
$X_{1},\cdots,X_{k},$ linearly independent over $\mathbb{R},$ such that
$\Gamma$ is the set of integer linear combinations of the $X_{k}$'s. (See
\cite[Exer. 3.18]{Wa} or \cite[Lem. 3.8]{BtD}.)

Now let $\frak{z}_{1}$ be the real span of $X_{1},\cdots,X_{k},$ and let
$\frak{z}_{2}$ be the orthogonal complement of $\frak{z}_{1}$ in $\frak{z},$
with respect to the fixed Ad-invariant inner product. Since $\frak{z}_{1}$ is
commutative, the image of $\frak{z}_{1}$ under the exponential mapping is
connected subgroup of $K,$ which is isomorphic to a torus, hence compact. Thus
the connected subgroup $H$ of $K$ whose Lie algebra is $\frak{k}_{1}%
+\frak{z}_{1}$ is a quotient of $K_{1}\times\left(  \frak{z}_{1}%
/\Gamma\right)  ,$ hence compact.

Next consider the map $\Psi:H\times\frak{z}_{2}\rightarrow K$ given by
\[
\Psi\left(  h,X\right)  =he^{X},
\]
which is a homomorphism because $\frak{z}_{2}$ is central. I claim that this
map is injective. To see this, suppose $\left(  h,X\right)  $ is in the
kernel. Then $h=e^{-X},$ which means that $h$ is in the center of $K,$ hence
in the center of $H.$ Now, $H$ is a quotient of $K_{1}\times\left(
\frak{z}_{1}/\Gamma\right)  ,$ so there exist $x\in K_{1}$ and $y\in\left(
\frak{z}_{1}/\Gamma\right)  $ such that $h=xy.$ Since $h$ is central and $y$
is central, $x$ is central as well. But the center of $K_{1}$ is finite, so
there exists $m$ such that $x^{m}=id.$ Since $y$ and $e^{X}$ are central, this
means that
\[
h^{m}=x^{m}y^{m}e^{mX}=y^{m}e^{mX}=id.
\]
But $y=e^{Y}$ for some $Y\in\frak{z}_{1},$ so we have $e^{mY}e^{mX}%
=e^{mY+mX}=id,$ which means that $mY+mX\in\Gamma.$ This means that $X=0,$
since $\frak{z}$ is the direct sum of the real span of $\Gamma$ and
$\frak{z}_{2},$ and so also $h=e^{-X}=id.$

Thus $\Psi$ is an injective homomorphism of $H\times Z_{2}$ into $K.$ The
associated Lie algebra homomorphism is clearly an isomorphism ($\frak{k}%
=\left(  \frak{k}_{1}+\frak{z}_{1}\right)  +\frak{z}_{2}$). It follows that
$\Psi$ is actually a diffeomorphism. To finish the argument, we need to show
that the Lie algebra of $H$ (namely, $\frak{k}_{1}+\frak{z}_{1}$) is
orthogonal to $\frak{z}_{2}.$ To see this, note that $\frak{k}_{1}$ and
$\frak{z}_{2}$ are automatically orthogonal with respect to any Ad-invariant
inner product (since the orthogonal projection of $\frak{k}_{1}$ onto
$\frak{z}_{2}$ is a Lie algebra homomorphism of a semisimple algebra into a
commutative algebra), and $\frak{z}_{1}$ and $\frak{z}_{2}$ are orthogonal
with respect to the chosen inner product, by the construction of $\frak{z}_{2}.$

\end{document}